\documentclass[]{article}
\usepackage[top=0.85in,left=0.75in,footskip=0.75in]{geometry}

\usepackage{changepage}
\usepackage[utf8]{inputenc}
\usepackage{textcomp,marvosym}
\usepackage{fixltx2e}
\usepackage{tikz}
\usepackage{subfigure}
\usetikzlibrary{positioning,shapes}
\usepackage{amsmath,amssymb}
\usepackage{cite}
\usepackage{nameref,hyperref}
\usepackage{microtype}
\usepackage{color}
\usepackage{bm}
\DisableLigatures[f]{encoding = *, family = * }
\usepackage{rotating}
\usepackage[aboveskip=1pt,labelfont=bf,labelsep=period,singlelinecheck=off]{caption}

\makeatletter
\renewcommand{\@biblabel}[1]{\quad#1.}
\makeatother

\usepackage{lastpage,fancyhdr,graphicx}
\usepackage{epstopdf}
\begin{document}

\begin{flushleft}
{\Large
\textbf \newline{
Tracking collective cell motion by topological data analysis }}

\bigskip
{\large L. L. Bonilla$^{1,2*}$, A. Carpio$^{2,3}$, C. Trenado$^1$}\\
\bigskip
{\bf 1 - G. Mill\'an Institute for Fluid Dynamics, Nanoscience \& Industrial Mathematics, and Department of Mathematics, Universidad Carlos III de Madrid, Avenida de la Universidad 30, 28911 Legan\'es, Spain}\\
{\bf 2 - Courant Institute of Mathematical Sciences, New York University, 251 Mercer St, New York, N.Y. 10012, USA}\\
{\bf 3 - Departamento de Matem\'atica Aplicada, Universidad Complutense, 28040 Madrid, Spain}
\bigskip

* Corresponding Author
\bigskip
\end{flushleft}

\section*{Abstract}
By modifying and calibrating an active vertex model to experiments, we have simulated numerically a confluent cellular monolayer spreading on an empty space and the collision of two monolayers of different cells in an antagonistic migration assay. Cells are subject to inertial forces and to active forces that try to align their velocities with those of neighboring ones. In agreement with experiments in the literature, the spreading test exhibits formation of fingers in the moving interfaces, there appear swirls in the velocity field, and the polar order parameter and the correlation and swirl lengths increase with time. Numerical simulations show that cells inside the tissue  have smaller area than those at the interface, which has been observed in recent experiments. In the antagonistic migration assay, a population of fluidlike Ras cells invades a population of wild type solidlike cells having shape parameters above and below the geometric critical value, respectively. Cell mixing or segregation depends on the junction tensions between different cells. We reproduce the experimentally observed antagonistic migration assays by assuming that a fraction of cells favor mixing, the others segregation, and that these cells are randomly distributed in space. To characterize and compare the structure of interfaces between cell types or of interfaces of spreading cellular monolayers in an automatic manner, we apply topological data analysis to experimental data and to results of our numerical simulations. We use time series of data generated by numerical simulations to automatically group, track and classify the advancing interfaces of cellular aggregates by means of bottleneck or Wasserstein distances of persistent homologies. These techniques of topological data analysis are scalable and could be used in studies involving large amounts of data. Besides applications to wound healing and metastatic cancer, these studies are relevant for tissue engineering, biological effects of materials, tissue and organ regeneration.

\section*{Author Summary}
Confluent motion of cells in tissues plays a crucial role in wound healing, tissue repair, development, morphogenesis and in numerous pathological processes such as tumor invasion and metastatic cancer. For such complex processes, controlled experiments help clarifying the roles of chemical, mechanical and biological cues. Among them, spreading of cellular tissues on an empty space and antagonistic migration assays between cancerous and normal cells are quite revealing. The interfaces between confluent cellular aggregates uncover properties thereof when a combination of modeling, numerical simulation and data analysis is used. Here we have modified an active vertex model with a dynamics that includes inertia, friction and active forces that tend to align cells based on interaction with its immediate neighborhood. Selecting appropriately junction tensions among cells and using the SAMoS software, we have succeed in simulating assays of cellular tissue spreading on an empty space and the invasion of healthy tissue by cancerous one. We have introduced topological data analysis to characterize, track  and compare in an automatic manner the interfaces of the tissue both in numerical simulations and from experimental data of normal and Ras modified precancerous Human Embryonic Kidney cells. We find good agreement when normal cells are solidlike and modified cells are liquidlike according to their shape parameters. In addition, cell variability means that a fraction of randomly distributed cells favor mixing, the others segregation. Topological data analysis techniques are scalable and could be used in studies involving  large amounts of data. Besides applications to wound healing and metastatic cancer, these studies are relevant in ascertaining how the biophysical features of materials may affect tissue and organ regeneration.


\section*{Introduction}\label{sec:1}
Confluent motion of epithelial cell monolayers  \cite{fri95,dur05,fri03,pou07,fri09,wei09,tre09,ror09,cat10,pet10,ang10,ang11,tre11,ror12,ber13,ser12,sep13,bru14,rav15,par15,ber16,por16,sme16,tay17,hak17,vol18,moi19,gan20} is crucial in many biological processes, such as morphogenesis \cite{fri09,vol18}, biological pattern formation \cite{cat10,sme16}, biological aggregation and swarming \cite{ber13,ber16}, tissue repair \cite{pou07,bru14,rav15}, development \cite{wei09}, and tumor invasion and metastasis \cite{fri95,fri03,fri09,gan20}. It serves as a relatively simple paradigm for collective motion of cells that retain their cell-cell junctions as they move on a two dimensional (2D) substrate. Confluent cellular motion can be tracked and visualized in experiments. Velocity and stress fields can be obtained by particle imaging velocimetry (PIV), time resolved cellular motion is observed using time-lapse imaging and fluorescence microscopy, traction microscopy allows to measure the forces that cells exert on the substrate as they move \cite{pou07,dur05,ser12}. Collective cell migration also poses challenging questions in soft and active matter physics, as it may exhibit fluid, solid or glass behavior with interesting flocking and jamming/unjamming transitions \cite{ang11,mar13,meh14,par15,bi16,mal17,gia17,gia18,pal19}. Interesting dynamics occurs as an epithelial cell aggregate advances through an empty space, as in wound healing, or it collides and encroaches a different  tissue, as in cancer invasion. Advancing cellular fronts may display wave phenomena \cite{ser12,rod17}, grow fingers \cite{oua09,sep13,ale19}, or breakdown and interpenetration against an oppositely moving front \cite{por16,moi19}. Different aspects of these phenomena have been studied by models ranging from macroscopic continuum mechanics to detailed subcellular agent models \cite{hak17,mar13,oua09,bar17,ale19model}.  

Here we combine particle dynamics \cite{sep13} with the active vertex model (AVM) \cite{bar17} to provide a cellular dynamics perspective on monolayers colliding in antagonistic migration assays (AMA) \cite{por16,moi19} or on monolayers spreading over an empty space \cite{oua09,ang10,sep13,ale19}. The resulting model describes the collective migration dynamics of a large number of cells and implements exchanges of neighboring cells automatically (T1 transitions) \cite{bar17}. In contrast to the usual overdamped dynamics in the AVM, the dynamics of the cell centers is underdamped. The underdamped AVM incorporates internal dissipation of cells through a friction parameter, a Vicsek-like velocity alignment of neighboring cells \cite{vic95,vic12,meh14}, noise and and active forces that may include cell polarity. We calibrate its parameters so that the simulations agree with experiments. Parameters for collective migration to an empty space are calibrated for Madin-Darby canine kidney (MDCK) cells \cite{pet10,sep13}. In AMA between MDCK cells, Ras modified cells collapse and are pushed backward by normal cells, which detect the former by an ephrin related mechanism \cite{por16}. The repulsive interactions between the two cell types drives cell segregation, produce sharp borders \cite{tay17}, and may generate deformation waves at the interface between the two cell types that propagate across the monolayers \cite{rod17}. In agreement with experiments in the literature, simulations of spreading test with our model exhibit formation of fingers in the moving interfaces, there appear swirls in the velocity field, and the polar order parameter and the correlation and swirl lengths increase with time, all of which has been observed in experiments \cite{oua09,pet10,ang10,sep13,ale19,pet11}. Our model is quite flexible, which gives it some advantages when describing behavior across different scales. Compared with particle models with underdamped dynamics, our model does not require introducing leader cells \cite{sep13} to account for fingering instabilities. Compared to continuum models \cite{ale19}, stochasticity enables our model to reproduce the observed spatial autocorrelation of the velocity \cite{ang10}. Simulations of our model show that cells in a finger of a moving interface may exhibit fast irregular oscillations in their velocity (periods of about one hour). This has been reported in early experiments \cite{pet11}. Our underdamped dynamics also predicts that cells inside an aggregate spreading onto an empty space have smaller area than those at the tissue interface. This prediction has been corroborated by experiments \cite{lv20}. Simulating the AVM with overdamped dynamics, we observe the opposite: cells at the interface and fingers have smaller are than cells inside the tissue \cite{bar17}.

In AMA with Human Embryonic Kidney (HEC) cell assemblies, precancerous Ras modified cells displace normal cells \cite{moi19}. These latter experiments have been interpreted using continuum mechanics in a simple biophysical model through phenomenological couplings \cite{ale19}, without recourse to biochemical signaling mechanisms and without clear relations to cellular processes. In this paper, we consider  wild type (wt) HEC cells to be solidlike whereas invading Ras cells are fluidlike and push the former backward. Experiments show that wt HEC cells keep their shape and area quite unchanged whereas Ras HEC cells may change shape and undergo larger deformations. This enforces our characterization of wt and Ras HEC cells as solidlike and fluidlike, respectively. As time elapses, there are cell exchanges and islands of one cell type form inside the tissue of the other cells, which characterizes a flocking liquid state \cite{mal17,gia18,ale19model}. In AMA with MDCK cells, the roles are inverted: Ras cells are solidlike and wt cells are fluidlike. The precise form of the separating interface among monolayers of different cell type depends on cell parameters governing segregation vs aggregation of these cells. We characterize it by topological data analysis (TDA). A measure of cellular diversity in the junction tensions produces islands of one type of cells inside the monolayer of the other cells, which is reflected in TDA of simulations and experiments. Cell cohesion given by the underdamped AVM, the cell alignment rule and the active noise force produce fingers in interfaces during assays of cell invasions of empty spaces rendering unnecessary to assume a different phenotype for lead cells \cite{sep13}. Recapitulating, our model explains a wide variety of experiments on confluent motion of cellular aggregates onto free space (wound healing) and invasion of one aggregate by another (antagonistic assays, cancer). It does this by choosing judiciously physical parameters such as cellular junction tension, adhesion and those in active forces. Fine tuning of parameters may require a deeper study of experimental data. One promising area where our results are very relevant is the study of the biophysical features of materials as they affect tissue and organ regeneration (materiobiology, tissue engineering) \cite{li17}. 

Recent experiments have connected metastasis in colorectal cancer to wound healing and tumor invasion of tissue using appropriate molecular markers \cite{gan20}. Thus, our description of spreading of cellular tissue and antagonistic migration assays using our modified active vertex model might be relevant for metastatic cancer. In particular, we have shown the role of cellular junction tensions in cell invasion, agglomeration and segregation. Promising mechanisms include Notch signaling pathways \cite{boa15} and models of the epithelial-mesenchymal transition and cancer stem cell formation \cite{boc17,boc18}. Incorporating these cellular mechanisms to our vertex model may pave the way to future progress in this area, much as incorporating the Notch signaling pathway to cellular Potts models helps understanding many aspects of angiogenesis \cite{veg20}.  Understanding precise biochemical mechanisms influencing cell-cell contact and confluent cellular tissue may help develop therapies for metastatic cancers \cite{boc18}. 

When studying spreading and collisions between cellular aggregates, the interfaces become rough and can shed and absorb groups of cells. It is important to be able to track automatically these changes for long time numerical  simulations and experiments generate large data sets that is hard to visualize and follow. For the first time in studies of confluent motion of cellular aggregates, we use  topological data analysis of time series generated by numerical simulations to automatically group, track and classify the advancing interfaces of cellular aggregates. Topological changes in the interfaces are reflected in barcodes and persistence diagrams of clusters and holes that change with scales (filtration parameters) \cite{car09,ede10} and themselves evolve in time. We track and study these changes by means of bottleneck or Wasserstein distances \cite{ede10,ker17}. Measuring these changes with time in data available from experiments and comparing with data from numerical simulations allows characterizing milestones in confluent motion of aggregates and the important time scales involved. In this work, we use techniques of topological data analysis with some data taken from experiments and a modest amount of data from numerical simulations so as to explain techniques and results in a clear manner. However, our techniques are scalable and could be used in studies involving large amounts of data, as,  for example, those generated to characterize zebrafish patterns by combining machine learning and topological data analysis  \cite{mcg20}.

The paper is organized as follows. The Section Model describes the models we simulate. The numerical values of the parameters are calibrated so as to reproduce experimental observations of collective cell migration in two different cases: an aggregate spreading to an empty space and the collision of two different cellular monolayers in antagonistic migration assays. The results of the numerical simulations are detailed in the Section of that name. We characterize the structure of advancing and interpenetrating cell fronts by means of topological data analysis in the Section Formation of Islands and Topological Data Analysis. Our conclusions are described in the last Section. The Appendix provides additional background on topological data analysis for the readers' ease of use, and it details our study of evolving interfaces of a spreading aggregate by taking slices of cells near the front.

\section*{Model}\label{sec:2}
We modify an active vertex model (AVM) \cite{bar17} and simulate it by adapting the SAMoS software \cite{samos}. The AVM combines the Vertex Model (VM) for confluent epithelial tissues \cite{hon80,mar13} with active matter dynamics \cite{bar17}. Sometimes what we call AVM following Ref.~\cite{bar17} is called {\em an active self-propelled Voronoi model} \cite{ale19model}. Let us describe first the VM, then the AVM and our modification of its dynamics.

\paragraph{Vertex Model.} The VM assumes that all cells in the epithelium are roughly the same height and thus that the entire system can be well approximated as a two-dimensional sheet. The conformation of the tissue in the VM is computed as a configuration that simultaneously optimizes area and perimeter of all cells. Two neighboring cells share a single edge, which is a straight line. Three junction lines typically meet at a vertex, although vertices with a higher number of contacts are also possible. The model tissue is therefore a mesh consisting of polygons (i.e., cells), edges (i.e., cell junctions), and vertices (i.e., meeting points of three or more cells). Each configuration of the mesh has the following associated energy 
\begin{eqnarray}
E_\text{VM} = \displaystyle\sum_{i=1}^N\!\left[\frac{K_i}{2}\, (A_i-A_i^0)^2 + \frac{\Gamma_i}{2} P_i^2\right]\! + \displaystyle\sum_{\langle\mu,\nu\rangle} \Lambda_{\mu\nu} \, l_{\mu\nu}. \label{eq1}
\end{eqnarray}
Here $N$ is the total number of cells, $A_i$ is the area of the cell $i$, $A^0_i$ is its reference area, and $K_i$ is the area modulus, i.e., a constant with units of energy per area squared measuring how hard it is to change the area of the cell. $P_i$ is the cell perimeter and $\Gamma_i$ (with units of energy per length squared) is the perimeter modulus that determines how hard it is to change perimeter $P_i$. $l_{\mu\nu}$ is the length of the junction between vertices $\mu$ and $\nu$, and $\Lambda_{\mu\nu}$ is the tension of that junction (with units of energy per length). The sum in the last term is over all pairs of vertices that share a junction. Note that the model allows for different cells to have different area and perimeter moduli as well as reference areas, allowing for modelling of tissues containing different cell types. The cell area and perimeter can be written in terms of vertex coordinates. Thus, vertex positions together with their connectivities uniquely determine the energy of the epithelial sheet. The main assumption of the VM is that the tissue will always be in a configuration which minimizes the total energy in Eq.~(\ref{eq1}). To implement the VM, we determine the positions of vertices that minimize $E_\text{VM}$ for a given set of parameters $K_i$, $\Gamma_i$, and $\Lambda_{\mu\nu}$. Cell rearrangements are modelled by introducing moves that change appropriately the connectivity among cells.

While the moduli $K_i$ and $\Gamma_i$ are positive, $\Lambda_{\mu\nu}<0$. When the cell $i$ shares junctions only with others of the same type, $\sum_{\langle\mu,\nu\rangle}\Lambda_{\mu\nu} l_{\mu\nu}=\Lambda_{\mu\nu}\sum_{\langle\mu,\nu\rangle}l_{\mu\nu}= \Lambda_{\mu\nu} P_i$, and this term can be put together with the perimeter term, thereby yielding $\frac{\Gamma_i}{2} (P_i-P_i^0)^2$ plus an unimportant constant, provided $P_i^0=-\Lambda_{\mu\nu}/\Gamma_i>0$. Thus the junction tension $\Lambda_{\mu\nu}$ determines the target perimeter of a type of cell. Let us assume that there are two cell types, 1 and 2, with moduli $K_j$, $\Gamma_j$, $j=1,2$, $\Lambda_{11}$, $\Lambda_{22}$, $\Lambda_{12}$, and target areas and perimeters $A_j^0$, $P_j^0$, $j=1,2$, respectively. We can complete squares and drop additive constants, thereby obtaining
\begin{eqnarray}
E_\text{VM} = \displaystyle\sum_{j=1}^2\sum_{i_j=1}^{N_j}\!\left[\frac{K_j}{2}\, (A_{i_j}-A_j^0)^2 + \frac{\Gamma_j}{2} (P_{i_j}-P_j^0)^2\right]\! + (2\Lambda_{12}-\Lambda_{11}-\Lambda_{22})\displaystyle\sum_{\langle\mu,\nu\rangle} l_{\mu\nu}, \label{eq2}
\end{eqnarray}
in which $N_1+N_2=N$. 

Clearly, $\Lambda_{12}<(\Lambda_{11}+\Lambda_{22})/2$ implies that energy is minimized when the number of junctions between both types of cells increases. Cells of different types therefore tend to mix. Conversely, when $\Lambda_{12}>(\Lambda_{11}+\Lambda_{22})/2$ cells of different type segregate, as suppressing junctions between cells of  different type minimizes energy. There is also a competition between the two first terms in Eq.~\eqref{eq2} to minimize energy. Assume $\Lambda_{12}=(\Lambda_{11}+\Lambda_{22})/2$ and therefore the last term in Eq.~\eqref{eq2} vanishes. The shape index $p_j^0=P_j^0/\sqrt{A_j^0}=|\Lambda_{jj}|/(\Gamma_j\sqrt{A_j^0})$ controls the ratio of the type $j$ cell perimeter to its area. For the VM, the value $p^{0*}=3.812$ (which corresponds to pentagons) separates solidlike and fluidlike behavior of the tissue \cite{bi16}. For $p^0<p^{0*}$, cortical tension is prevalent over cell-cell adhesion, cells do not exchange neighbors and the monolayer is solidlike. For $p^0>p^{0*}$, cell-cell adhesion dominates, neighbor exchanges occur, and the cellular tissue behaves like a fluid \cite{bi16}.

\begin{figure}[h]
\begin{center}
\includegraphics[width=9cm]{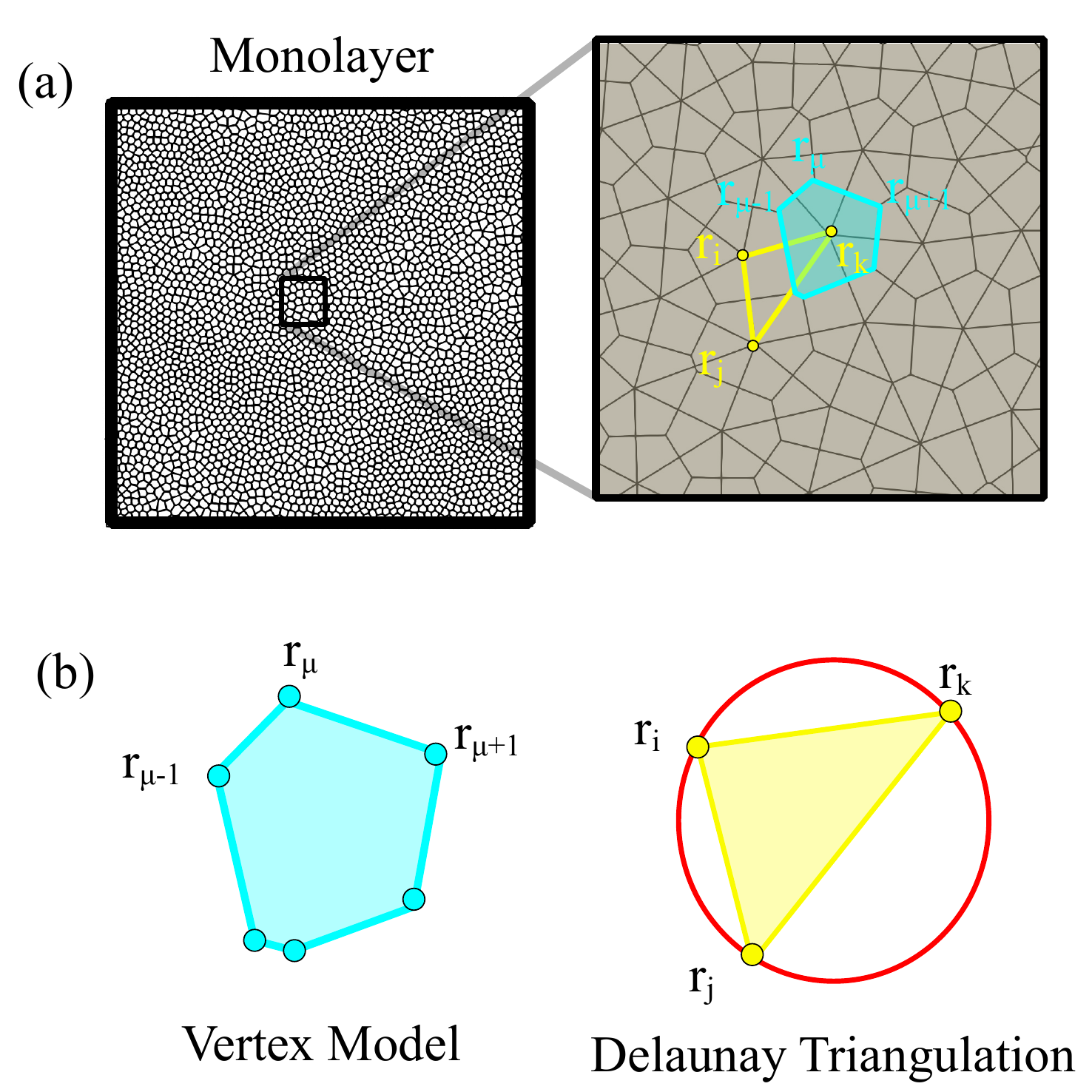}
\end{center}
\caption{Voronoi tessellation and Delaunay triangulation. (a) Here $\mathbf{r}_\mu$ are vertices of polygons in the Voronoi tessselation and $\mathbf{r}_i$ are centers of polygons that are vertices of Delaunay triangles. Here the zoom of a monolayer shows (b) the Voronoi tesselation and the Delaunay triangulation. \label{fig1}}
\end{figure}

\paragraph{Active Vertex Model.} To introduce dynamics in the VM, we have to go from polygon vertices $\mathbf{r}_\mu$ to polygon centers that represent cells, $\mathbf{r}_i$, consider these centers as particles and introduce dynamics for them \cite{bar17}. In this, the AVM is similar to the self-propelled Voronoi model \cite{bi16}. The core assumption of the AVM is that the tissue configurations that optimize the energy in Eq.~\eqref{eq1} correspond to the Voronoi tessellations of the plane with polygons as cells and cell centers acting as Voronoi seeds. Given a Voronoi tessellation, we consider its dual Delaunay triangulation, comprising Voronoi seeds and the edges joining them (triangles), which have the property that no seed is inside the circumcircle of any triangle; see Fig.~\ref{fig1}. From a Voronoi tesselation it is straightforward to obtain the dual Delaunay triangulation and vice versa. However, working with Delaunay triangulations has an advantage: they retain their nature when triangle vertices move by flipping edges conveniently \cite{bar17}, whereas Voronoi tessellations do not. The latter have to be reset after motion of polygon vertices. In the AVM, the area $A_i$ in Eq.~\eqref{eq1} of the cell $i$ is the area of the associated Voronoi polygon, $\Omega_i$, given by the following discrete version of Green's formula:
\begin{eqnarray}
A_i= \frac{1}{2}\sum_{\mu\in\Omega_i} (\mathbf{r}_\mu\times\mathbf{r}_{\mu+1})\cdot \mathbf{N}_i, \label{eq3}
\end{eqnarray}
where $\mathbf{r}_\mu$ is the position of vertex $\mu$, and $\mathbf{N}_i$ is a unit vector perpendicular to the surface of the polygon. For the 2D tissue $\mathbf{N}_i$ is directed along the $z$ axis and therefore does not depend on the position of the vertices. The sum in Eq.~\eqref{eq3} is over all vertices of the Voronoi cell and we close the loop with $\mu+1=1$ when $\mu$ equals the total number of vertices in the cell, $N_{\Omega_i}$. The cell perimeter is
\begin{eqnarray}
P_i= \frac{1}{2}\sum_{\mu\in\Omega_i} |\mathbf{r}_\mu-\mathbf{r}_{\mu+1}|. \label{eq4}
\end{eqnarray}
The relation between the vertices $\mathbf{r}_\mu$ of the Voronoi polygons (i.e., cells) and the vertices $\mathbf{r}_i$ of the Delaunay triangles (seeds of the Voronoi polygons, i.e., cell centers) is 
\begin{equation}
\mathbf{r}_{\mu}=\frac{\lambda_{1}\mathbf{r}_{i}+\lambda_{2}\mathbf{r}_{j}+\lambda_{3} \mathbf{r}_{k}}{\lambda_{1} + \lambda_{2}+\lambda_{3}}. \label{eq5}
\end{equation}
Here $\mathbf{r}_{i}$, $\mathbf{r}_{j}$ and $\mathbf{r}_{k}$ are position vectors of the corners of the triangle and $\lambda_{i}$, $i=1,2,3$, are the barycentric coordinates; cf. Fig.~\ref{fig1}, and Ref.~\cite{bar17} for details. The forces $\mathbf{F}_i =-\nabla_{\mathbf{r}_i} [E_\text{VM}+V_\text{soft}(|\mathbf{r}_i-\mathbf{r}_j|)]$ are \cite{bar17}
\begin{align}
\mathbf{F}_{i} & =-\sum_{k=1}^{N}\frac{K_{k}}{2}\left(A_{k}-A_{k}^{0}\right)\sum_{\nu\in\Omega_{k}}\left[\mathbf{r}_{\nu+1,\nu-1}\times\mathbf{N}_{k}\right]^{T}\left[\frac{\partial\mathbf{r}_{\nu}}{\partial\mathbf{r}_{i}}\right]\nonumber \\
 & -\sum_{k=1}^{N}\Gamma_{k}P_{k}\sum_{\nu\in\Omega_{k}}\left(\hat{\mathbf{r}}_{\nu,\nu-1}-\hat{\mathbf{r}}_{\nu+1,\nu}\right)^{T}\left[\frac{\partial\mathbf{r}_{\nu}}{\partial\mathbf{r}_{i}}\right]\nonumber \\
 & -\sum_{k=1}^{N}\sum_{\nu\in\Omega_{k}}\left[\Lambda_{\nu-1,\nu}\hat{\mathbf{r}}_{\nu,\nu-1}-\Lambda_{\nu,\nu+1}\hat{\mathbf{r}}_{\nu+1,\nu}\right]^{T}\left[\frac{\partial\mathbf{r}_{\nu}}{\partial\mathbf{r}_{i}}\right]\!\nonumber\\
 &+ k\sum_{\langle j,i\rangle} (2a-|\mathbf{r}_i-\mathbf{r}_j|)\,\frac{\mathbf{r}_i-\mathbf{r}_j}{|\mathbf{r}_i-\mathbf{r}_j|}\,\Theta(2a-|\mathbf{r}_i-\mathbf{r}_j|).\label{eq6}
\end{align}
Here $\left[\frac{\partial\mathbf{r}_{\nu}}{\partial\mathbf{r}_{i}}\right]$ is the $3\times3$ Jacobian matrix connecting coordinates of cell centres with coordinates of the dual Voronoi tessellation, and the non-commutative row-matrix product $\left[\cdot\right]^{T}\left[\cdot\right]$ is a $3\times1$ column vector. $\Theta(x)=1$ if $x>0$, else $\Theta(x)=0$, is the Heaviside unit step function. We have included a range repulsive force of short range $a$ that avoids self intersections of the triangulation for very obtuse triangles \cite{bar17}.

In the AVM, the usual dynamics for the cell centers is a gradient flow of the energy in Eq.~\eqref{eq1}, that is overdamped dynamics with forces $\mathbf{F}_i$ given by Eq.~\eqref{eq6}, plus active forces $f_a\mathbf{n}_i$, and stochastic forces $\bm{\nu}_i$ \cite{bar17}
\begin{eqnarray}
\gamma\,\dot{\mathbf{r}}_i= f_a\mathbf{n}_i+\mathbf{F}_i+\bm{\nu}_i, \quad\gamma_r\dot{\theta}_i=\bm{\tau}_i \cdot \mathbf{N}_i+\nu_i^r(t), \label{eq7}
\end{eqnarray}
where $\dot{\mathbf{r}}_i=d\mathbf{r}_i/dt$, $\bm{\tau}_i$ is the torque acting on the polarity $\mathbf{n}_i=(\cos\theta_i,\sin\theta_i)$, $\mathbf{N}_i$ is the local normal to the cell surface (the unit length vector in the $z$-direction), $\gamma_r$ is the orientational friction, and $\nu_i^r(t)$ is a zero mean Gaussian white noise responsible for orientational randomness, such that $\langle\nu_i^r(t)\nu_j^r(t')\rangle=2D_r\delta_{ij}\delta(t-t')$. Terms aligning cell velocity or shape to polarity or terms aligning the polarity of different cells can be included in the energy of Eq.~\eqref{eq1} \cite{bar17}. A particularly simple dynamics follows from $f_a=v_0$ (constant active force), $\bm{\nu}_i=\bm{\tau}_i=\bm{0}$ \cite{bi16}. The AVM describes naturally cell motion and accounts for patterns of the confluent tissue observed on multiple scales, from cell sizes to much larger distances. Furthermore, cell contacts are generated dynamically from the positions of cell centers.

\paragraph{Dynamics including velocity alignment and inertia.} In this work, we shall modify the AVM dynamics. Instead of Eq.~\eqref{eq7}, we shall use the particle dynamics of Ref.~\cite{sep13} but with different forces between particles. As discussed in Ref.~\cite{sel05}, trajectories of motile cells can be explained by assuming that their acceleration is a certain functional of velocity. Despite the mass of the cell being so small that inertia is negligible compared with typical forces exerted on the cell, the formula for acceleration resembles Newton's second law \cite{sel05}. In this formula, a linear damping term represents dissipative processes coming from friction with substrate, with other cells, or rupture of adhesion bonds. Active memory terms, which are linear in the velocity, may propel single cells and account for the observed non-monotonic velocity autocorrelation \cite{sel05}. When considering cellular tissue, Sep\'ulveda {\em et al} model cells as actively motile particles and replace the memory terms by Vicsek-like alignment ``forces'' \cite{vic95}, and interparticle and random ``forces'' \cite{sep13}. Thus, the acceleration in these models is a consequence of the collective motion of cells and the interaction with the environment and it does not follow from Newton's second law with a mass given by that of a single cell. However, we will continue denoting by {\em forces} (per unit mass) the terms comprising the acceleration \cite{sep13}. In contrast to Eq.~\eqref{eq7}, the cells in Ref.~\cite{sep13} are not self-propelled, so that they can stop their motion and start moving again if there are missing cells in their neighborhood and the active force is zero:
\begin{eqnarray}
\dot{\mathbf{r}}_i=\mathbf{v}_i,\quad\dot{\mathbf{v}}_i = -\alpha\mathbf{v}_i + \sum_{\langle j,i\rangle}\!\left[\frac{\beta}{n_i}(\mathbf{v}_j-\mathbf{v}_i)+\mathbf{f}_{ij}\right]\!+\boldsymbol{\varphi}_i + \sigma_0\boldsymbol{\eta}_i(t),\quad\tau \dot{\bm{\eta}}_i=-\bm{\eta}_i+ \bm{\xi}_i(t).\quad
\label{eq8}
\end{eqnarray}
Here, the sum is over the nearest neighbors of the vertex $i$ of the Delaunay triangulation, $n_i$ is the number of these neighbors, the friction coefficient $\alpha$ comes from internal cell friction or adhesion to the substrate or other cells. The term containing the coefficient $\beta$ tries to synchronize the velocity of the nearest neighbor cells that of the $i$th cell and it is similar to the Vicsek model \cite{meh14,vic95,vic12,bon19}. $\mathbf{f}_{ij}$ is the force per unit mass exerted by cell $j$ on cell $i$. In our simulations we use $\sum_{\langle j,i\rangle}\mathbf{f}_{ij}=\mathbf{F}_i/m_i$, where $\mathbf{F}_i$ is given by Eq.~\eqref{eq6}, and not by an interparticle potential as in Ref.~\cite{sep13}. $m_i$ is a reference mass, for example $m_i={\gamma \gamma_r^2 / D_r}.$ The active forces are $\boldsymbol{\varphi}_i +\sigma_0 \boldsymbol{\eta}_i(t)$. In Ref.~\cite{sep13}, $\boldsymbol{\varphi}_i=0$ and $\boldsymbol{\eta}_i(t)$ is a zero mean Ornstein-Uhlenbeck noise, representing a stochastic force with nonzero correlation time $\tau$. $\bm{\xi}_i(t)$ is a zero-mean delta-correlated Gaussian white noise. For spreading tests, we have used the numerical values of the parameters indicated in Table \ref{t1}. For antagonistic migration assays, we have used the numerical values collected in Table \ref{t2}. As we shall see in the description of the numerical simulations, the dynamics given by Eq.~\eqref{eq8} with our choice of forces allows us to reproduce many features observed in experiments.

\begin{table}[ht]
\centering
\begin{tabular}{cccccccccc}
	$\alpha$ & $\beta$ &  $\tau$ & $\sigma_0$ & $K$& $\Gamma$ & $\Lambda$& $\lambda$& $l_0$ & $\zeta$\\ \hline
h$^{-1}$	&  h$^{-1}$    & h  & $\frac{\mu\text{m}}{\text{h}^2}$ & - & - & - & - & -& -\\
0.534 & 41.36 & 0.56 &95 & 1 & 0.1 & -1& 0.1& 0 & 0.5\\ 
\end{tabular} 
\caption{Parameters corresponding to the experiments with MDCK cells in Ref.~\cite{sep13}.}
\label{t1}
\end{table}

\begin{table}[ht]
\begin{center}
\begin{tabular}{ccccccc}
	$\alpha$ & $\beta$ &  $\tau$ & $\sigma_0$ & $K_j$ & $\Gamma_j$& Figure \# \\ \hline
h$^{-1}$	&  h$^{-1}$    & h  & $\frac{\mu\text{m}}{\text{h}^2}$ & - & -& -\\
0.0602 & 13.85 & 1.66 &55.88 & 1 & 1& 8\\ 
0.42 & 0.602 & 1.66 &13.97 & 1 & 1& 9, 10 \\
\end{tabular}
\end{center}
\caption{Two sets of parameters corresponding to the experiments with HEK cells in Ref.~\cite{moi19}.}
\label{t2}
\end{table}

\begin{figure}[h]
\begin{center}
\includegraphics[width=12cm]{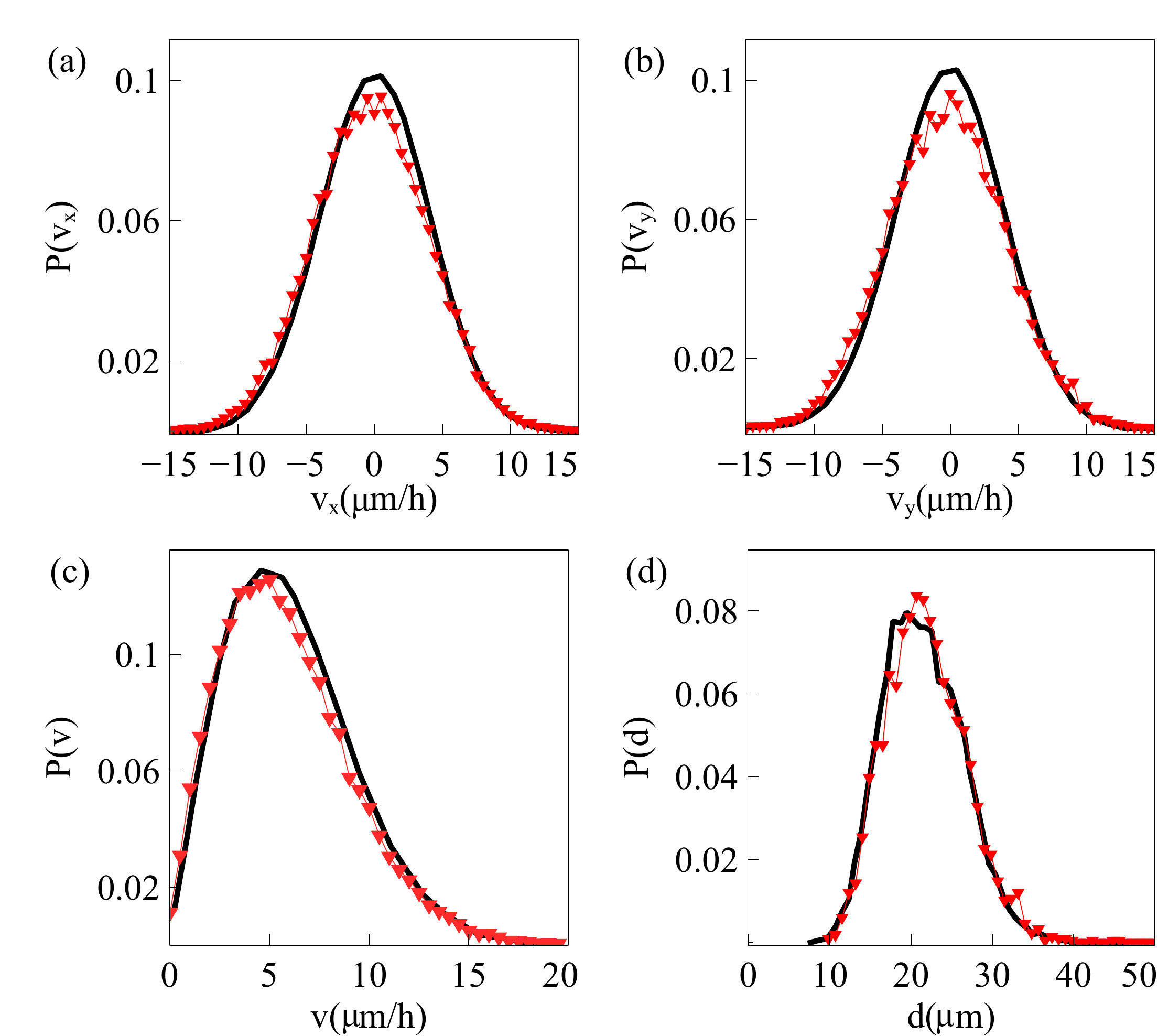}
\end{center}
\caption{Probability distribution function (PDF) for particle velocities: (a) $v_x$, (b) $v_y$, (c) $v=|\mathbf{v}|$; and (d) mean distance $d$ between neighboring particles; after the initialization procedure (red triangles) as compared to the experimentally observed PDF (black line) \cite{sep13}. Parameter values are those in Table \ref{t1}.\label{fig2}}
\end{figure}

\begin{figure}[!h]
\begin{center}
\includegraphics[width=12cm]{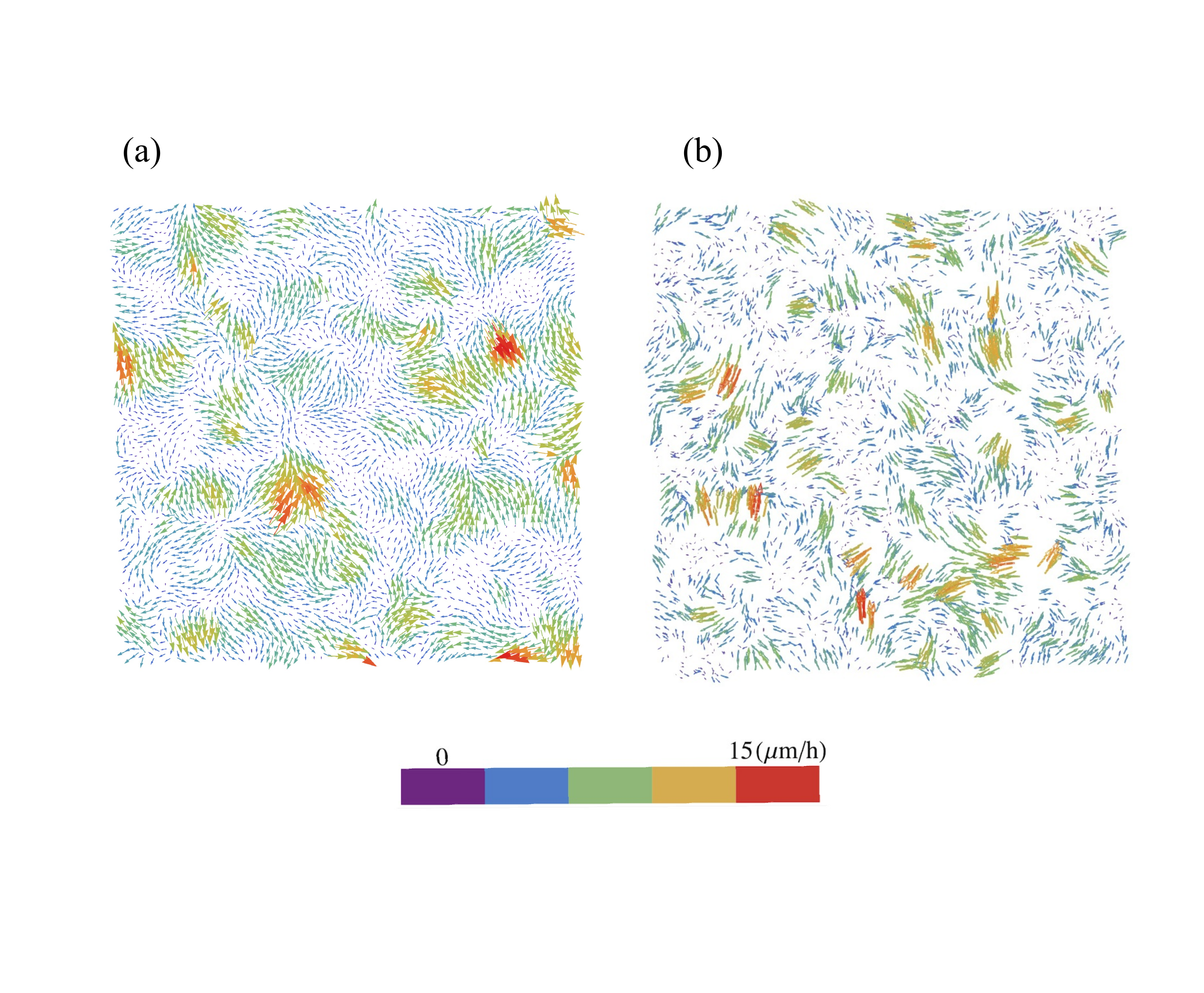}
\end{center}
\caption{Velocity field obtained from (a) experiments \cite{sep13}, (b) simulations after the initialization procedure. Parameter values are those in Table \ref{t1}.\label{fig3}}
\end{figure}

\begin{figure}[!h]
\begin{center}
\includegraphics[width=10cm]{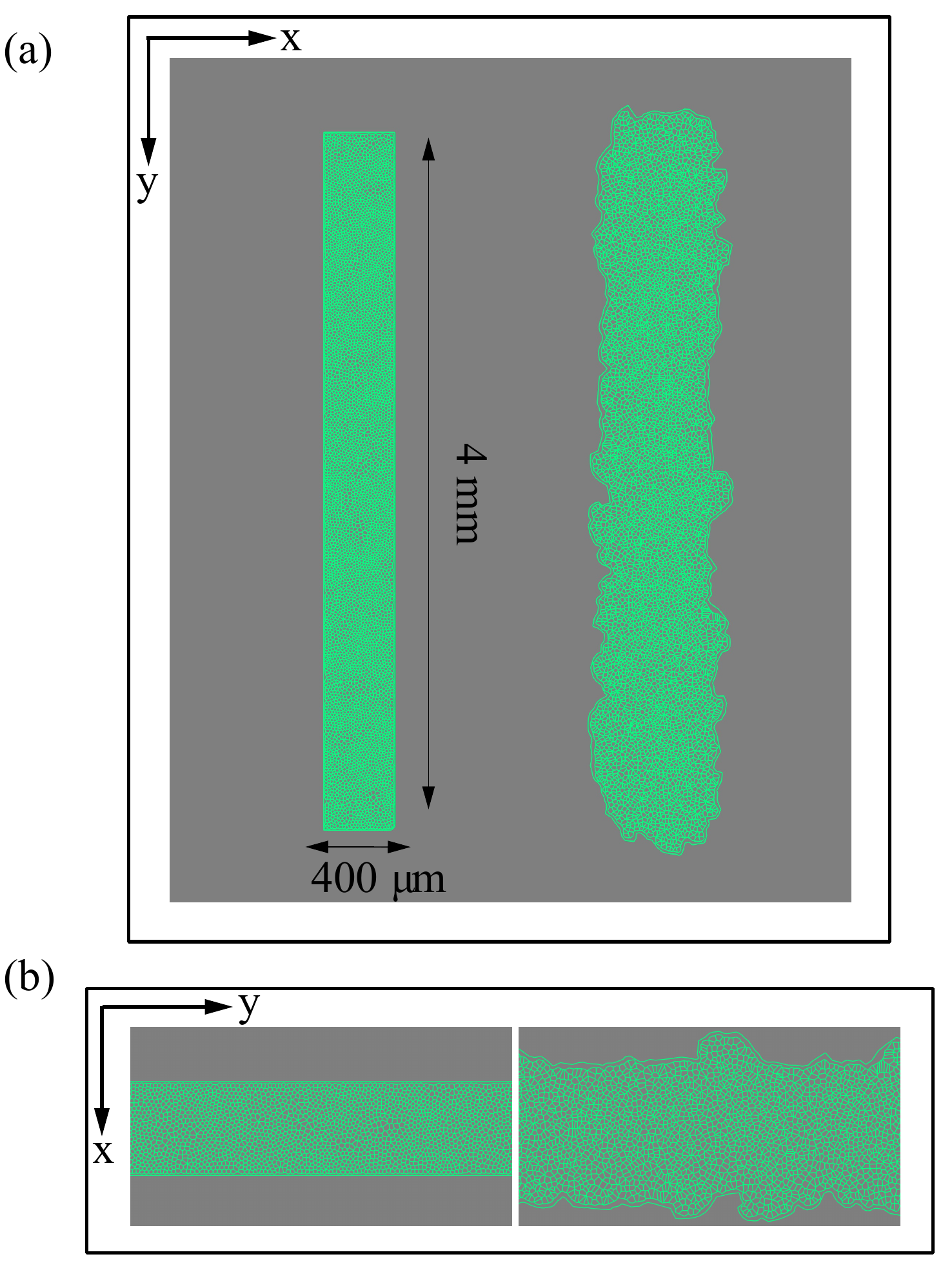}
\end{center}
\caption{Initial configuration and configuration after $20$ h of stencil removal showing the formation of fingers according to the numerical simulation of the model. (a) Full view, (b) zoom.  Initial box size is 1.6 mm$^2$, $P^0=10$, $A^0=\pi$, and shape index $p^0=5.65$. Parameter values are those in Table \ref{t1}. \label{fig4} }
\end{figure}

\begin{figure}[!h]
\begin{center}
\includegraphics[width=14cm]{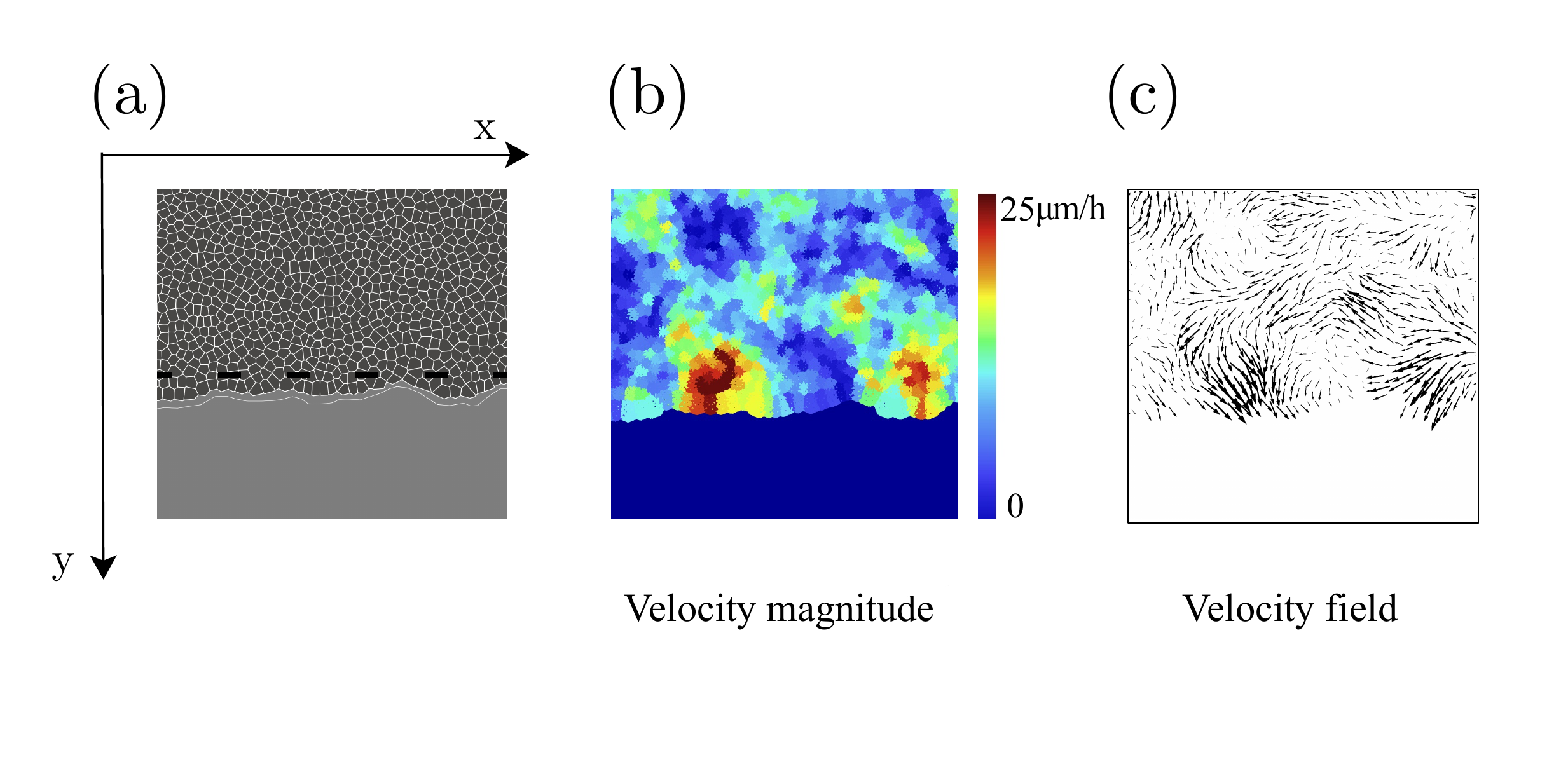}
\end{center}
\caption{Cell velocity field after $2$ h of stencil removal in an invasion configuration calculated from simulations of the model with the parameters of Fig.~\ref{fig4} and Table \ref{t1}. (a) Phase contrast visualizing cells, (b) profile of cell speed (modulus of velocity), (c) velocity field. These panels should be compared with those obtained from experimental MDCK data in Fig.~2A of Ref.~\cite{pet10}.} \label{fig5} 
\end{figure}

\paragraph{Boundaries.} The cells at the boundary between a cellular monolayer and an empty space, or between tissues, are special. They may form actin cables, thereby having a line tension and a bending stiffness \cite{bar17}:
\begin{eqnarray}
&&E_\text{lt}=\frac{1}{2}\sum_{\langle i,j\rangle}\lambda_{ij}(l_{ij}-l_0)^2, \label{eq9}\\
&&E_\text{bend}=\frac{1}{2}\sum_i\zeta_i(\theta_{i}-\pi)^2,\quad \cos\theta_i=\frac{\mathbf{r}_{ji}\cdot\mathbf{r}_{ki}}{|\mathbf{r}_{ji}|\, |\mathbf{r}_{ki}|}. \label{eq10}
\end{eqnarray}
Here the modulus $\lambda_{ij}$ is the line tension of the edge connecting vertices $i$ and $j$, $l_{ij}=|\mathbf{r}_{ij}|$ ($\mathbf{r}_{ij}=\mathbf{r}_i-\mathbf{r}_j$) is the edge length (of preferred magnitude $l_0$), $\zeta_i$ is the bending stiffness of angle $\theta_i$ at the boundary particle $i$, and $\mathbf{r}_j$ and $\mathbf{r}_k$ are the positions of boundary particles to the left and to the right of particle $i$. The line tension energy of Eq.~\eqref{eq9}  tries to keep boundary edges at a length $l_0$ whereas the bending energy of Eq.~\eqref{eq10} tries to keep the boundary line flat. The sums in these formulas are over boundary particles only and we assume that each boundary cell has exactly two boundary neighbors \cite{bar17}. 

\paragraph{Initial condition.} A random configuration of the particles comprising a confluent cell monolayer is usually different from those configurations observed in experiments. Thus, we have to carry out an initialization stage until the particle configuration is compatible with their observed velocity distributions. For spread tests, we proceed as follows. We set a square box of size 1 mm$^ 2$ area, $N \approx 4000$ particles  (comparable to the number of cells in the experiments), the packing ratio and the particle mean velocity. Then, we numerically solve Eqs.~\eqref{eq1} and \eqref{eq8} with forces $\boldsymbol{\varphi}_i =\mathbf{0}$ and $\sum_{\langle j,i\rangle}\mathbf{f}_{ij}=\mathbf{F}_i/m_i$, $\mathbf{F}_i$ given by Eq.~\eqref{eq6}, until the velocity probability density functions (PDFs) of the experiments are fitted. The parameters adjusted to the experimental data at early time (30 min after stencil removal) are listed in Table \ref{t1}. We stop the initialization stage when the distribution of mean distances between particles is close to the initial distribution as observed in experiments and displayed in Fig.~\ref{fig2}. From this simulation, we obtain the particle positions $\mathbf{r}_i$ and solve the underdamped AVM with forces given by Eq.~\eqref{eq6} and initial random directions for the particle velocities. As we can see in Fig.~\ref{fig3}, the velocity field obtained from the simulations, Fig.~\ref{fig3}(b), is very similar to that measured by PIV analysis \cite{sep13}, Fig.~\ref{fig3}(a). 

For AMA, we choose a random configuration having the same number of wt and Ras cells separated by a vertical straight line and we set a known velocity distribution from experiments \cite{moi19}. This represents the situation of the two monolayers when they first make contact. See details in the next section.

\section*{Results of numerical simulations}\label{sec:3}
We have simulated two different tissue configurations: (a) a cellular monolayer spreads over an  empty space, and (b) two monolayers comprising wild type and modified cells collide. In each case, the simulations are compared to relevant experimental observations.

\subsection*{Results for the spreading configuration.} Inspired by wound healing phenomena and experiments on tissue scratching, we are interested in the movement of an epithelium which encroaches on a virgin substrate. The experimental protocol consists of microfabricated stencils whose removal increases the motility of the epithelium. In our simulations, we consider a narrow strip configuration as that in Fig.~\ref{fig4}(a), which is similar to those in Ref.~\cite{pet11}. We adapt the SAMoS code \cite{samos} to simulate the AVM with dynamics given by Eqs.~\eqref{eq8} and \eqref{eq6}, in which $\boldsymbol{\varphi}_i = \mathbf{0}$. Parameter values are those in Table \ref{t1}. Cells migrate on the surface maintaining their junctions with their neighbors, which is enforced by the term proportional to $\beta$ in Eq.~\eqref{eq8}. During healing, noisy forcing in Eq.~\eqref{eq8} makes some cells to move faster that the others while keeping their contacts. This is the origin of the fingers or instabilities of the interface with the cell free space, which are illustrated by Fig.~\ref{fig4}(b), see Video 1 in supplemental material for the complete time evolution. In addition, cells on the interface, or close to it, may grow beyond the target area $A^0$ in Eq.~\eqref{eq1}. As they do so, each cell has a probability to divide into two daughter cells, which equals  $r_d\, (A-A^0)\, dt$. Here $dt$ is the time step and $r_d$ is the division rate. We have normalized the target area to $A^0=\pi$, $dt=0.05$, $r_d=0.01$, and we check whether the cell divides with ten times the above probability every 10 time steps that we observe $A>A^0$. With these parameters, there is some cell division near the interface of the confluent layer and the empty space.

\begin{figure}[!h]
\begin{center}
\includegraphics[width=13cm]{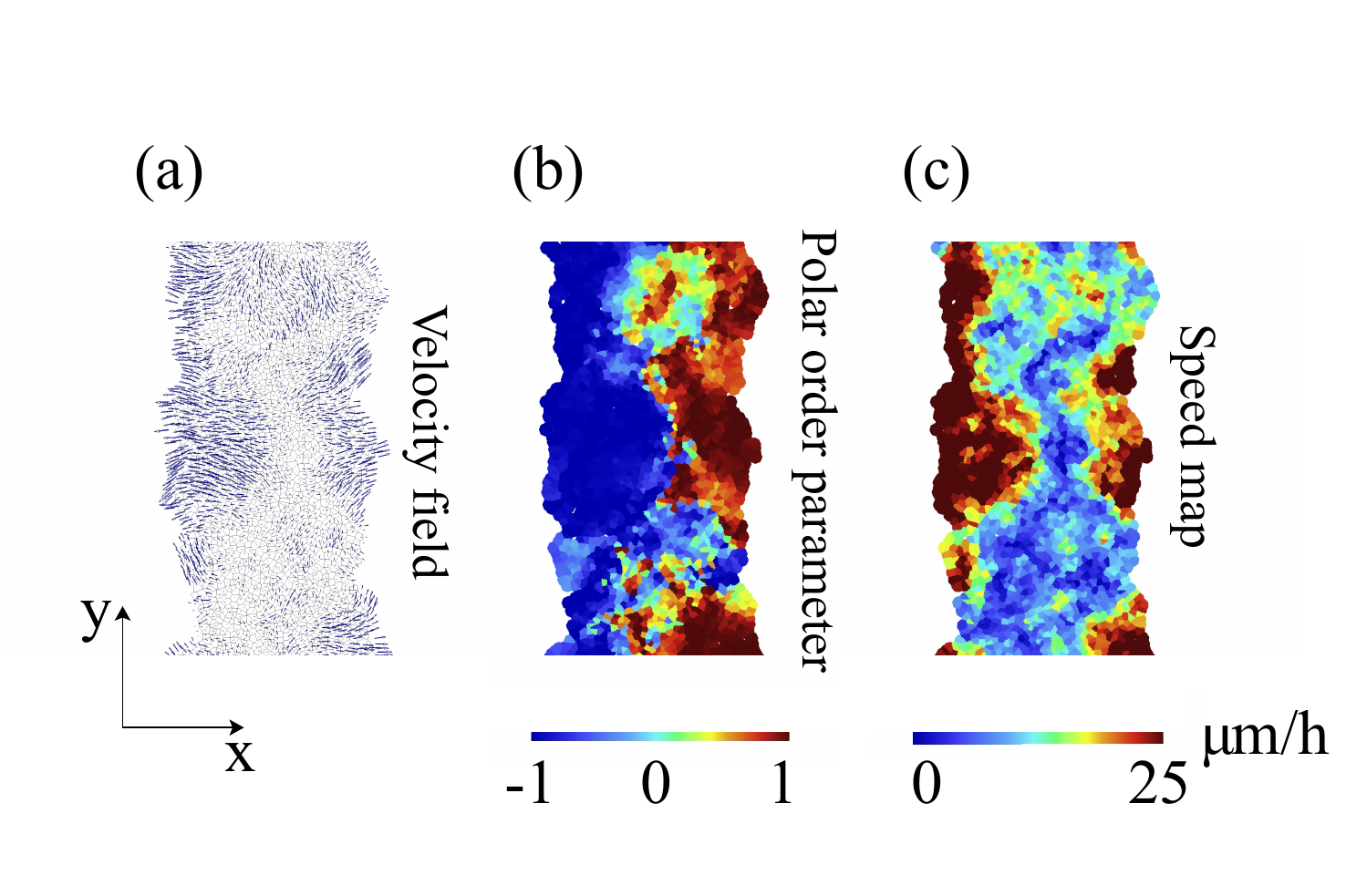}
\end{center}
\caption{(a) Numerically simulated cell velocity field, (b) local polar order parameter $\cos\vartheta_i$, and (c) speed ($|\mathbf{v}|$) map after $35$ h of stencil removal in an invasion configuration for a $400$ $\mu$m wide strip. Parameter values as in Fig.~\ref{fig5}. \label{fig6} }
\end{figure}

\begin{figure}[!h]
\begin{center}
\includegraphics[width=10cm]{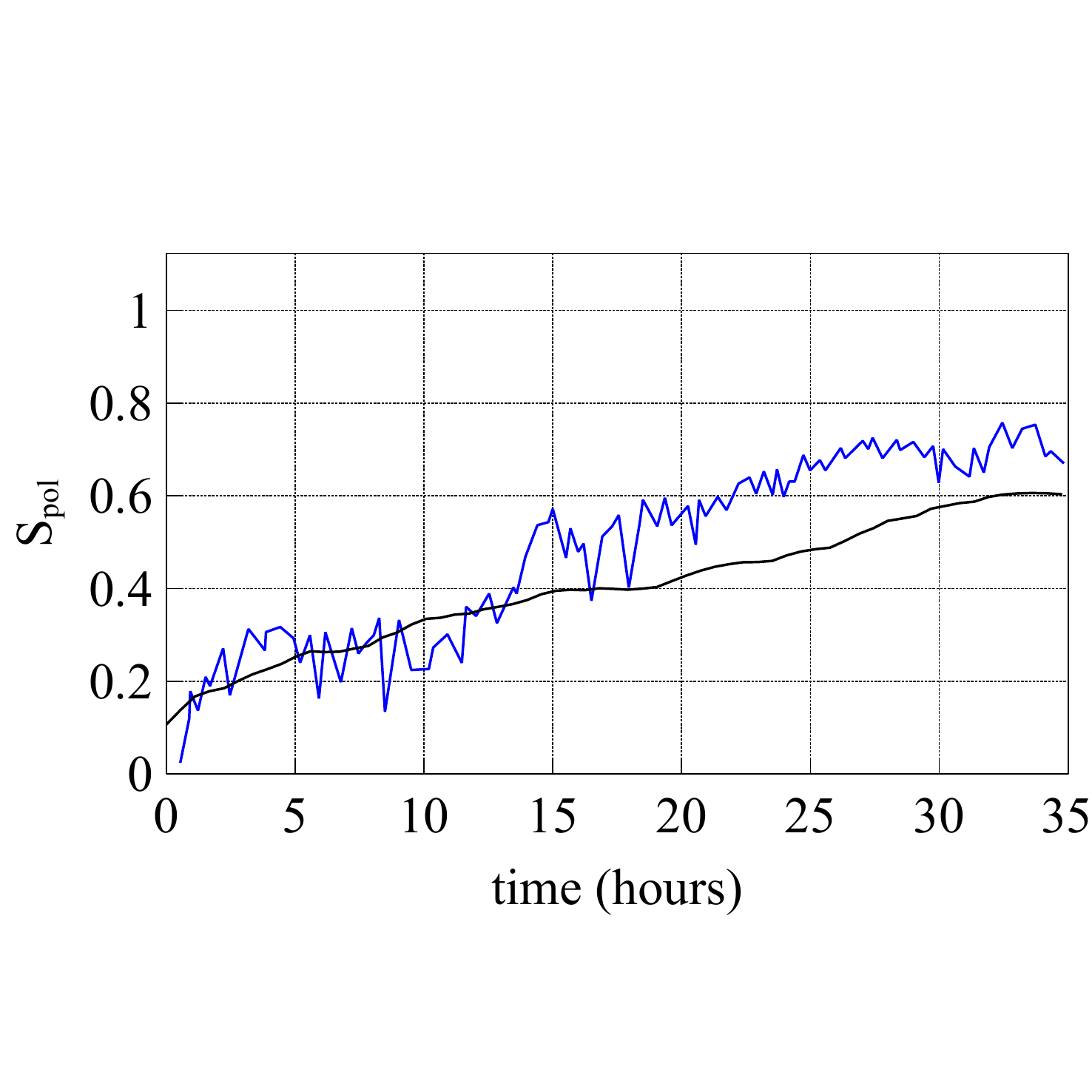}
\end{center}
\vskip-1cm
\caption{Evolution of the polar order parameter $S_\text{pol}(t)$ corresponding to Fig.~\ref{fig5}. Here $t=0$ corresponds to $1.5$ h after stencil removal \cite{pet10}. An average over $5$ simulations exhibits the same trend as measurements reported in Ref.~\cite{pet10} (jagged line). \label{fig7} }
\end{figure}

\begin{figure}[!h]
\begin{center}
\includegraphics[width=14cm]{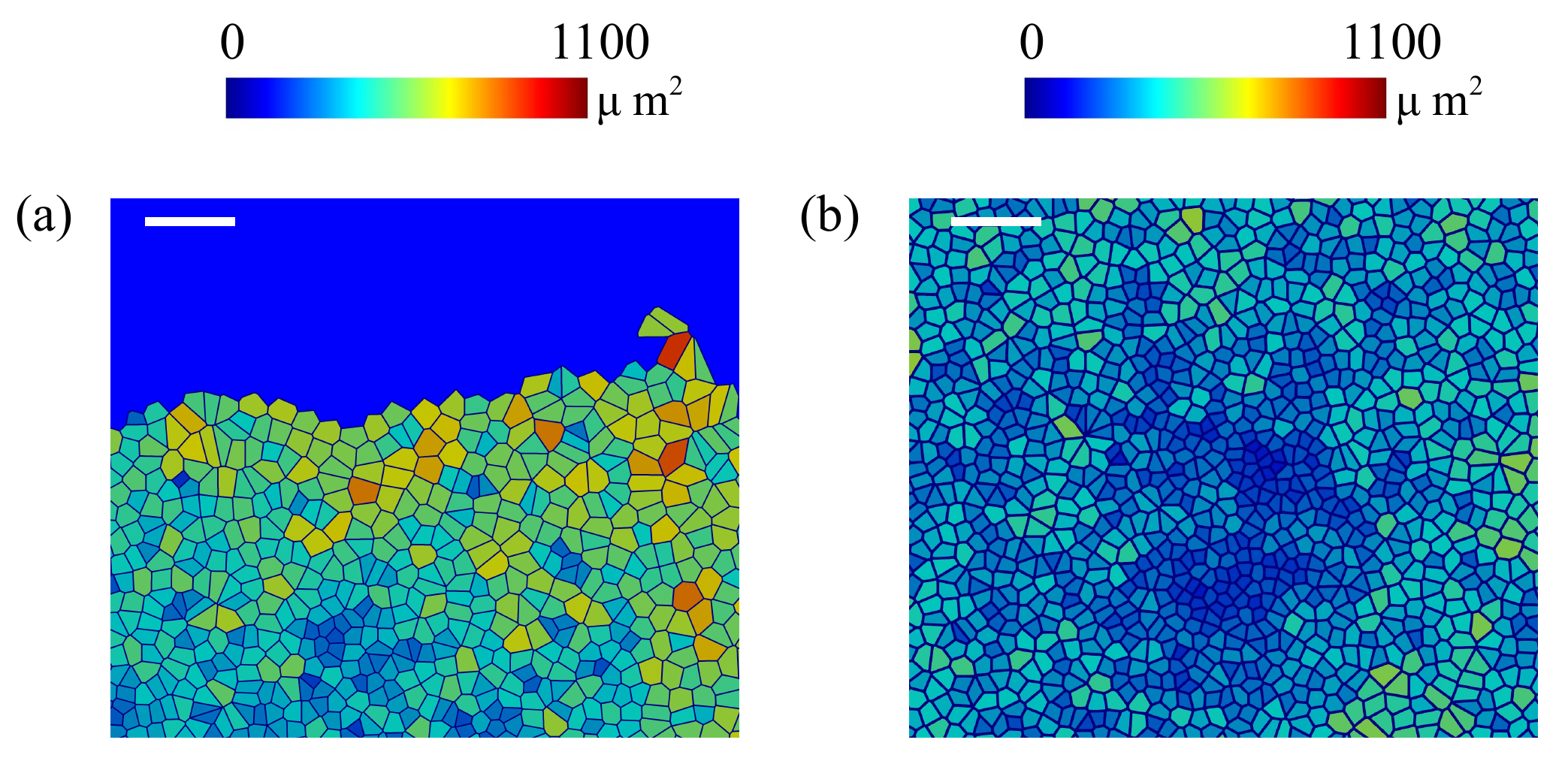}
\end{center}
\caption{Areas of cells during a simulation of a spreading configuration: (a) Area of cells near the interface, (b) area of cells far from the interface. Our simulations exhibits the same trend as measurements reported in Ref.~\cite{lv20}.}
\label{fig8} 
\end{figure}

\begin{figure}[!h]
\begin{center}
\includegraphics[width=12cm]{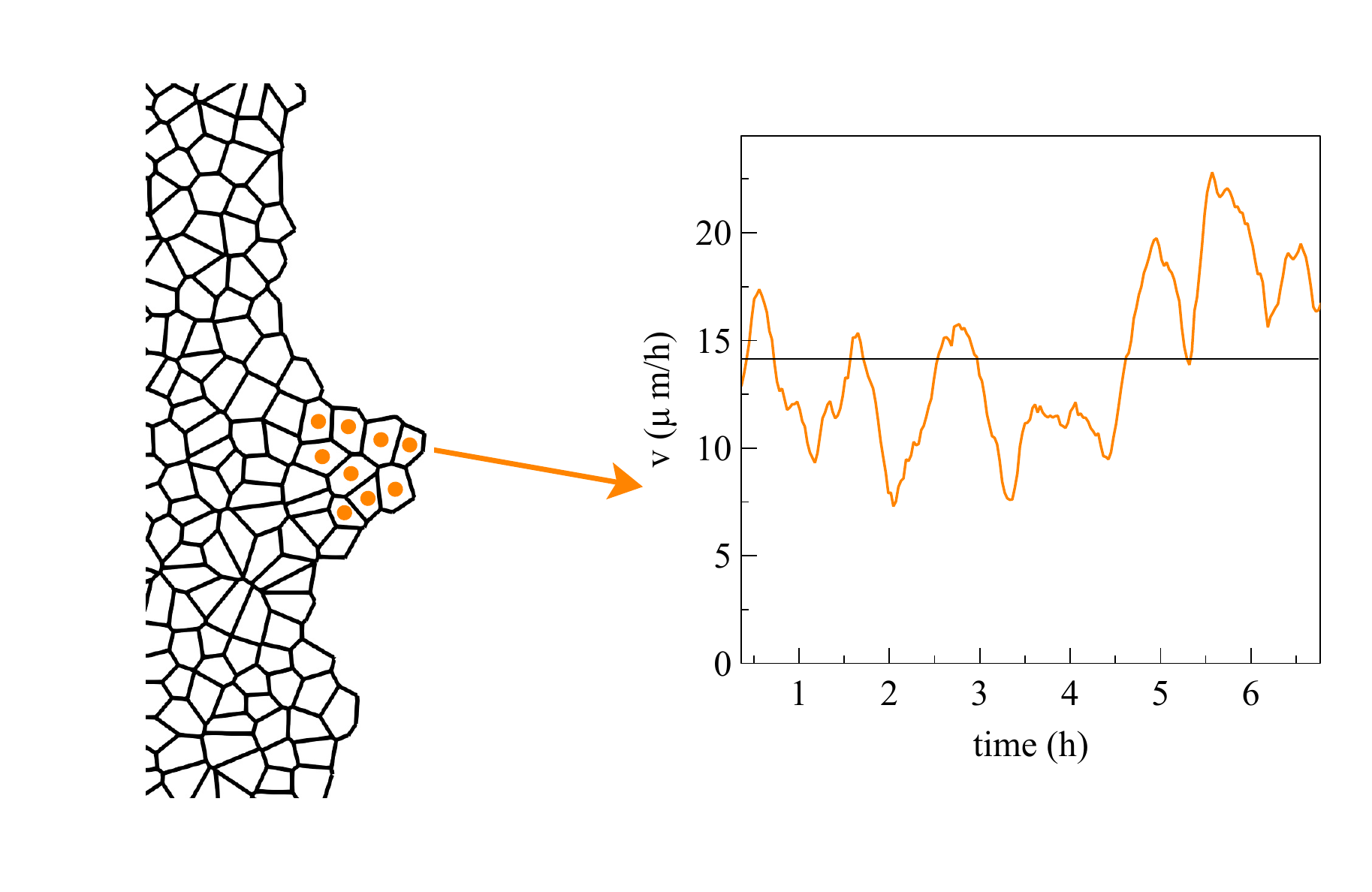}
\end{center}
\caption{Average velocity of the marked cells during finger expansion. The velocity of each cell oscillates in a similar but somewhat more irregular manner (not shown). \label{fig9} }
\end{figure}

\begin{figure}[!h]
\begin{center}
\includegraphics[width=12cm]{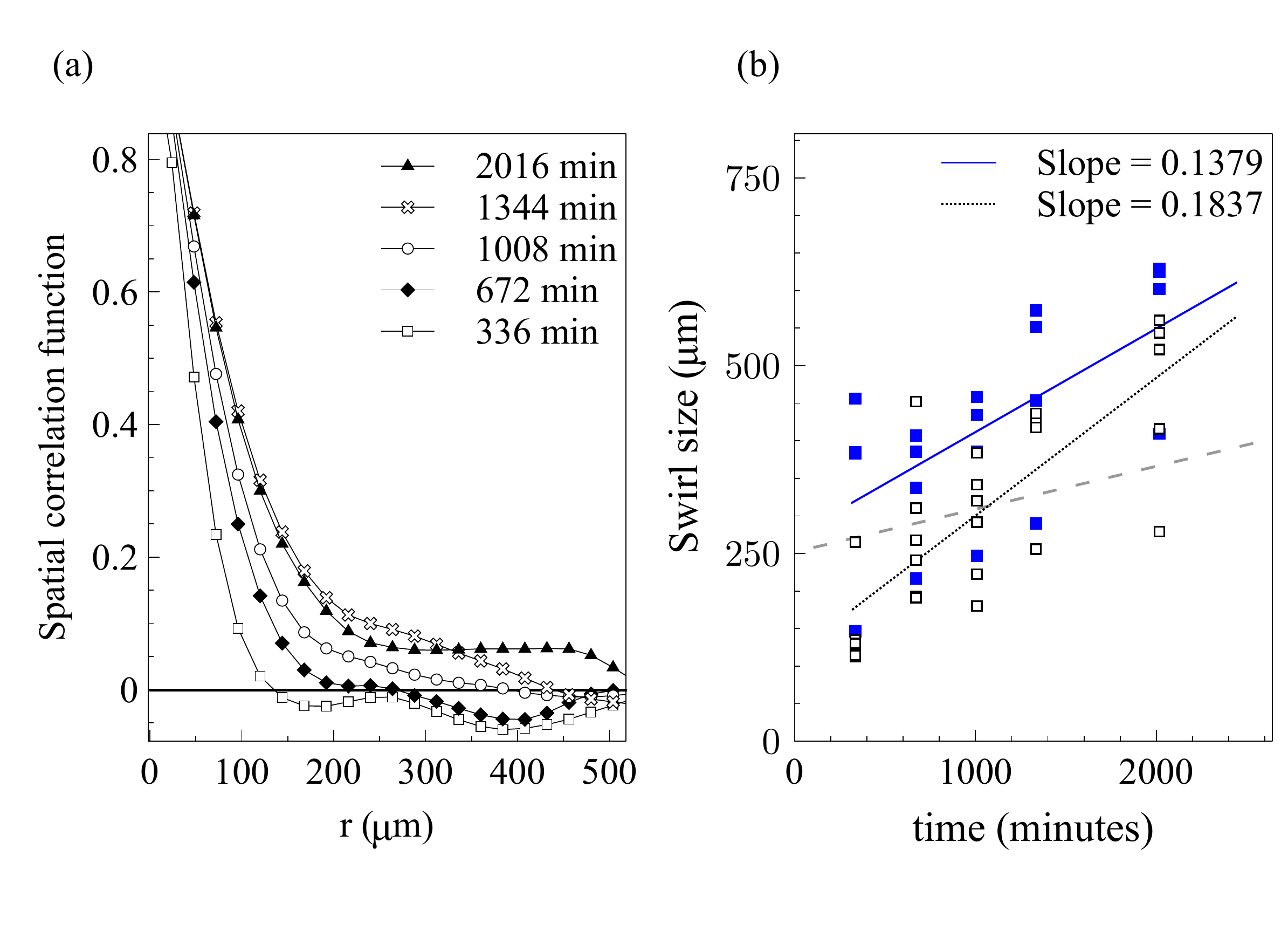}
\end{center}
\caption{(a) Spatial correlation function $I(r,t)$ corresponding to Fig.~\ref{fig6}(a) for different times. (b) Correlation length given by the first zero of $I(r)$ (empty squares) and swirl size given by the first local minimum of $I(r)$ (blue squares). Dashed line from swirl sizes in Ref.~\cite{ang10}. \label{fig10} }
\end{figure}

After two hours of stencil removal, the PIV recorded from the experiments reveals the complex movements that can appear inside the bulk of the tissue, cf. Fig.~2A of Ref.~\cite{pet10}. Cells do not move independently and their velocities are correlated. The presence of these cellular flows shows the existence of motion inside the monolayer \cite{ang10}. Similar to experiments, our simulations in Fig.~\ref{fig5} show that incipient fingers appear in areas of high speed; see also Video 2 in the supplemental material. In our simulations, we find these areas without having to postulate the existence of special leader cells. Having calculated numerically the velocity field, we can quantify the orientational motion inside the epithelium. Take for example, the configuration after $35$ h of stencil removal is shown in Fig.~\ref{fig6}. In addition to the velocity field and the speed (modulus of the velocity vector) map, we have depicted a density map of the polar order parameter $S_\text{pol}$, 
\begin{equation}
S_\text{pol}=\frac{1}{N}\sum_{i=1}^N\cos\vartheta_i,\quad \cos\vartheta_i= \frac{v_x(i)}{\sqrt{v_x(i)^2+v_y(i)^2}}.  \label{eq11}
\end{equation}
Here $\vartheta_i$ is the angle that the velocity vector of the $i$th cell forms with the outer normal to the strip (the $x$ axis in Fig.~\ref{fig4}). Fig.~\ref{fig6}(b) depicts the density plot of the cellular polar order parameter, $\cos\vartheta_i$, after $35$ h of stencil removal (similar to experimental data reported in Fig. 92 of Ref.~\cite{pet11}). Fig.~\ref{fig7} shows that an ensemble average of the polar order parameter (over 5 realizations, smooth line) increases with time and follows the same trend as the measurements reported in Ref.~\cite{pet10} (jagged line).  At early times, $S_\text{pol}$ in Fig.~\ref{fig7} does not exhibit a particular trend. The angles are distributed homogeneously and are not located in specific areas. After a while, the cells start orient themselves perpendicular to the strip, specially at the edges of the tissue, as shown in Fig.~\ref{fig6}. This effect occurs in strips of width larger than $300$ $\mu$m. On shorter strips, their two sides are no longer independent and the appearance of a finger changes the motion of the whole strip.

The fronts of advancing cells in Figs.~\ref{fig5} and \ref{fig6} clearly show the formation of fingers. The AVM keeps cells together while the term proportional to $\beta$ in Eq.~\eqref{eq8} induces a common average direction in their motion. This effect becomes stronger the larger $\beta$ is, which promotes and enforces finger formation. Thus, unlike the particle model of Ref.~\cite{sep13}, we do not need a longer range attractive potential interaction between cells. We do not need to distinguish leader cells to trigger finger formation \cite{sep13} because advancing cells at the forefront of the monolayer pull those behind them. A comparison of our simulation results in Figs.~\ref{fig5} and \ref{fig6} to the experiments reported in Refs.~\cite{pet10,pet11} shows that the appearance and size of the cell velocity field are  reproduced qualitatively. Our simulations show that the area and velocity of cells both increase as their distance to the boundary of the cellular tissue decreases. Fig.~\ref{fig8} shows that cells near the interface in a spreading configuration have larger areas than cells far from the interface. This is particularly noticeable in the fingers: the cells in them are faster and have a larger area than the cells elsewhere. The cells far from the tissue border are compressed and have smaller area than boundary ones. This prediction of the underdamped AVM with dynamics as in Eq.~\eqref{eq8} has been observed in experiments; see Fig.~4 of Ref.~\cite{lv20}. In experiments, the area of finger cells reaches larger values than in the simulations, which is related to the fact that we use a fixed target area for all cells and the cell area cannot depart arbitrarily far from target in the AVM. We have also simulated the AVM with the overdamped dynamics of Eq.~\eqref{eq7} and with the same boundary and initial conditions. In this case, after fingers are formed as in Fig.~6 of Ref.~\cite{bar17}, the interior cells far from the interface have larger area than cells at the boundary and in the fingers. This is also shown in Fig.~8 of Ref.~\cite{bar17}.  However, this behavior is contrary to experimental observations \cite{lv20}.

Our numerical simulations of spreading configurations show that the cells inside a finger move faster than those at other portions of the interface. We have observed that the average velocity of finger cells may oscillate irregularly about some average value with a short period of about one hour. Fig.~\ref{fig9} shows the average velocity of 9 finger cells during a 7 hour time interval. The velocity of a single cell in the finger oscillates somewhat more irregularly in a similar fashion. For much longer time intervals, the average velocity may experience an overall upward trend. The average velocity of boundary cells in flat regions also oscillates with time but it does not show a definite behavior over long time intervals: it may even display a downward trend. In experiments, the velocity of cells leading interfacial fingers has also been observed to oscillate rapidly and irregularly with periods of about one hour or less, which is similar to the findings based on numerical simulations of our model; see Fig.~101A of Ref.~\cite{pet11}. Some models based on continuum mechanics predict longer periods of tens of hours \cite{pet11}.

The velocity field in Fig.~\ref{fig6}(a) exhibits swirl patterns \cite{ang10}. To characterize them, we have depicted in Fig.~\ref{fig10}(a) the correlation function for the $x$-component of the velocity field:
\begin{equation}
I(|\mathbf{r}|,t)=\frac{\langle v_x^*(\mathbf{r}',t)\, v_x^*(\mathbf{r}'+\mathbf{r},t)\rangle_{\mathbf{r'}}}{\sqrt{\langle v_x^*(\mathbf{r}',t)^2\rangle_{\mathbf{r'}}\langle v_x^*(\mathbf{r}'+\mathbf{r},t)^2\rangle}_{\mathbf{r'}}}, \quad v_x^*(\mathbf{r},t)=v_x(\mathbf{r},t)-\langle v_x(\mathbf{r},t)\rangle_{\mathbf{r}}.\label{eq12}
\end{equation}
Here the averages are spatial averages over ${\mathbf{r'}}$ and also ensemble averages over simulations with different initial conditions. Fig.~\ref{fig10}(b) depicts the correlation length defined by the first zero of the correlation function and the swirl size defined by its first local minimum. Empty and blue squares correspond to values given by different simulations. The best fits to straight lines are also shown and compared to a similar line for Angelini {\em et al}'s experimental data \cite{ang10}. Clearly correlation length and swirl size increase with time, indicating that cells feel each other on increasingly larger regions as time elapses. This has been observed in other experiments and simulations \cite{pet10,moi19}. The correlation lengths given by our simulations agree quite well with values reported in the literature for similar observation times \cite{ang10,pet10,moi19}.

\subsection*{Results for the collision configuration.} 
\begin{figure}[h!]
\begin{center}
\includegraphics[width=12cm]{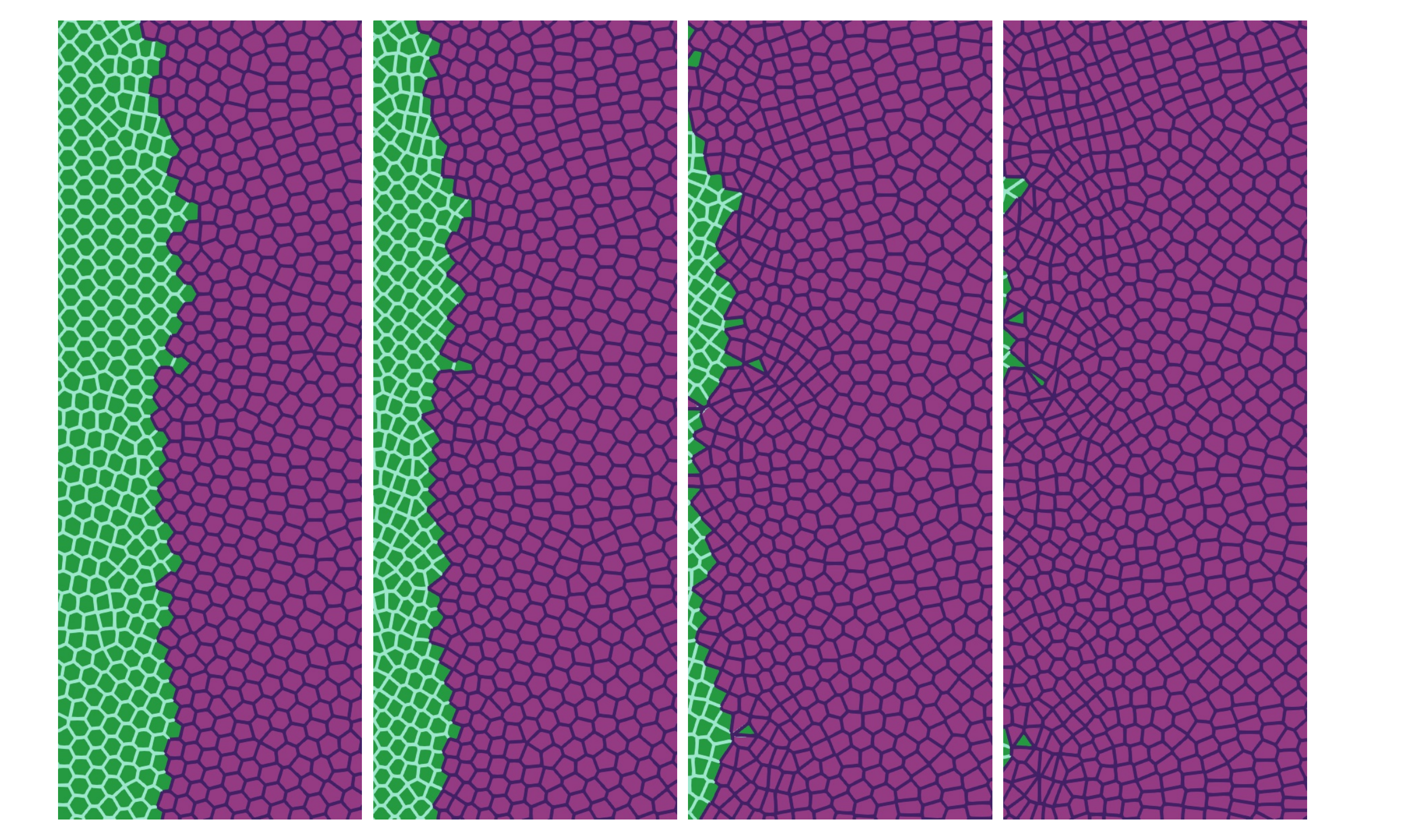}
\end{center}
\caption{Simulation of the antagonistic migration assay: one population advances pushing back the other. Junction tensions are $\Lambda_{11}=-6.2$, $\Lambda_{22}=-6.8$, which yield shape indices 3.50 (green cells) and 3.84 (magenta cells), respectively. Other parameters are listed in the first row of Table \ref{t2}, and $\Lambda_{12} =-7.0<\frac{1}{2}(\Lambda_{11}+\Lambda_{22})$ correspond to weak population mixing. Snapshots are taken at times 2 h, 6.5 h, 13 h, 20 h. 
\label{fig11} }
\end{figure}

Recently, Moitrier {\em et al} have reported confrontation assays between antagonistically migrating cell sheets \cite{moi19}. In their experiment, the two confluent cellular monolayers (wild type and modified Ras HEK cells) advance toward an intermediate empty space, collide and the Ras monolayer displaces the wt one. The experiment shows that the velocities of the cells decay exponentially fast the farther they are from the advancing fronts \cite{moi19}. If $x=L(t)$ is the position of the monolayer front, the velocity of the cells at position $x<L$ is $V^\text{wt}\exp[(x-L)/\lambda^\text{wt}]$ for the wt and $-V^\text{Ras}\exp[-(x-L)/\lambda^\text{Ras}]$ for the Ras cells at $x>L$. After the collision, these velocity functions remain the same but now $V^\text{wt}$ and $V^\text{Ras}$ acquire a common and lower value $-V^\text{interface}$. Moitrier {\em et al} interpret their experiments by comparing with simple solutions of a 1D continuum model \cite{moi19}. In our simulations, we use the SAMoS code to simulate the underdamped AVM cellular model with dynamics given by Eq.~\eqref{eq8}.  The invading Ras cells (magenta) move to the left whereas the wt cells (green) are pushed backward because they experience aversion to mixing with Ras cells. We model this situation by adding a negative active force $\boldsymbol{\varphi}^\text{Ras}_i =a^\text{Ras}\exp[-[x-L(0)]/\lambda^\text{Ras}]$ to Ras cells in Eq.~\eqref{eq8} for $x>L(0)$ (not included in Ref.~\cite{sep13}), whereas wt and Ras cells do not experience an active force if $x<L(0)$. We use $\lambda^\text{Ras}=410\,\mu$m, $a^\text{Ras}=9\,\mu$m/h$^2$, $L(0)=0$.  The active force $\boldsymbol{\varphi}^\text{Ras}$ keeps Ras cells moving to the left and pushing wt ones. Therefore we no longer need the synchronization force proportional to $\beta$ to keep cells moving in the same direction. Fig.~\ref{fig11} shows finger formation for the active force $\boldsymbol{\varphi}^\text{Ras}$ and for $\beta=13.85$ h$^{-1}$, which is smaller than the value in Table \ref{t1}. Other parameters are as indicated in Table \ref{t2}. 

Our underdamped AVM uses more features of wt and Ras cells obtained from the experiments than kept by continuum models. The latter lose features at distances close to the cell size. Continuum models fit friction, viscosity and strength of active forces for the two cell populations to explain how Ras cells invade the wt monolayer \cite{moi19}. 

\begin{figure}[h!]
\begin{center}
\includegraphics[width=12cm]{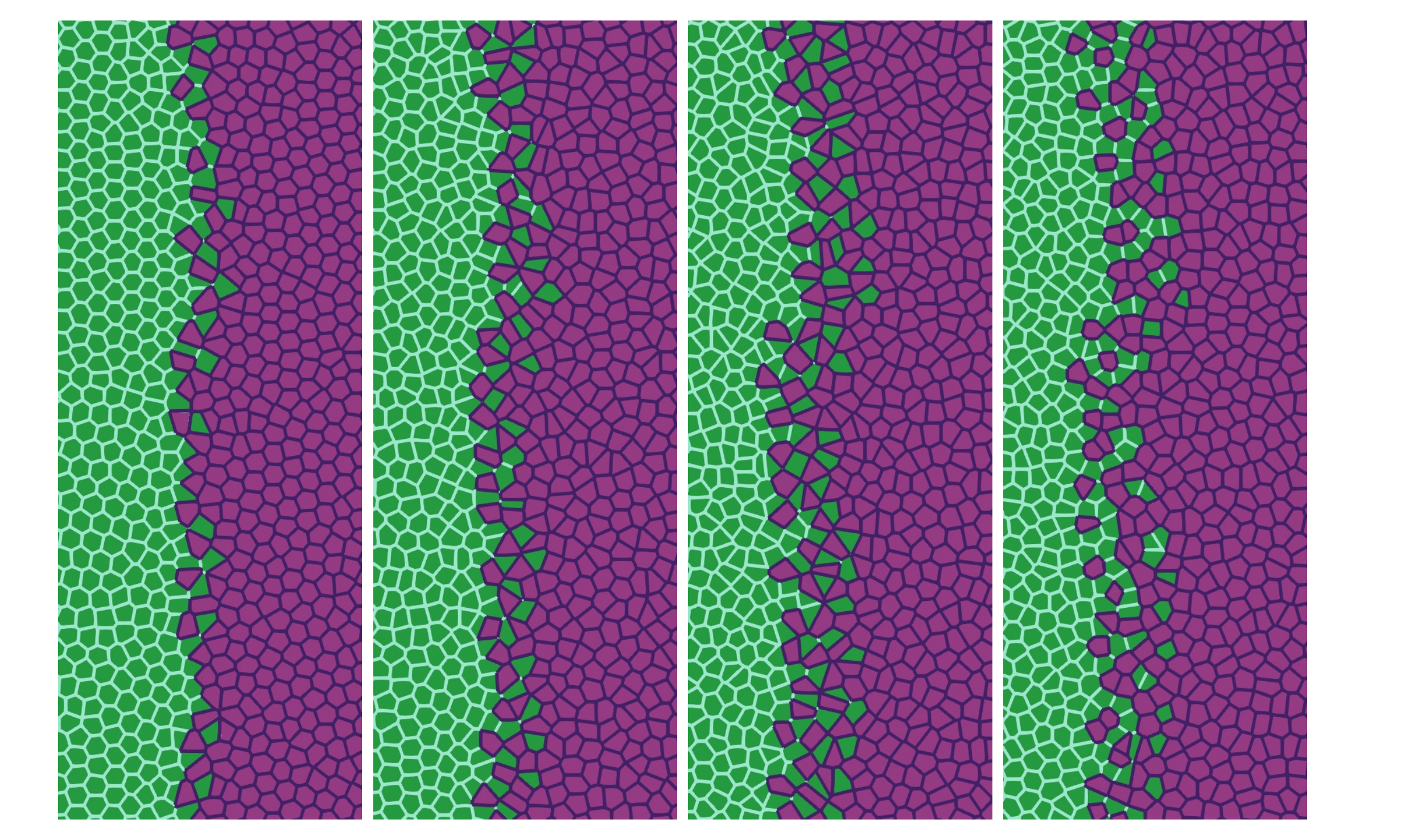}
\end{center}
\caption{Simulation of the antagonistic migration assay: creation of a extremely rugged interface. First and second snapshots: $\Lambda_{12}=-7.5<\frac{1}{2}(\Lambda_{11}+\Lambda_{22})$ (population mixing); third and fourth snapshots: $\Lambda_{12}=-6.0>\frac{1}{2}(\Lambda_{11}+\Lambda_{22})$ (population segregation). Other parameters are listed in the second row of Table \ref{t2} whereas times are as in Fig.~\ref{fig11}.  
\label{fig12} }
\end{figure}

\begin{figure}[h!]
\begin{center}
\includegraphics[width=12cm]{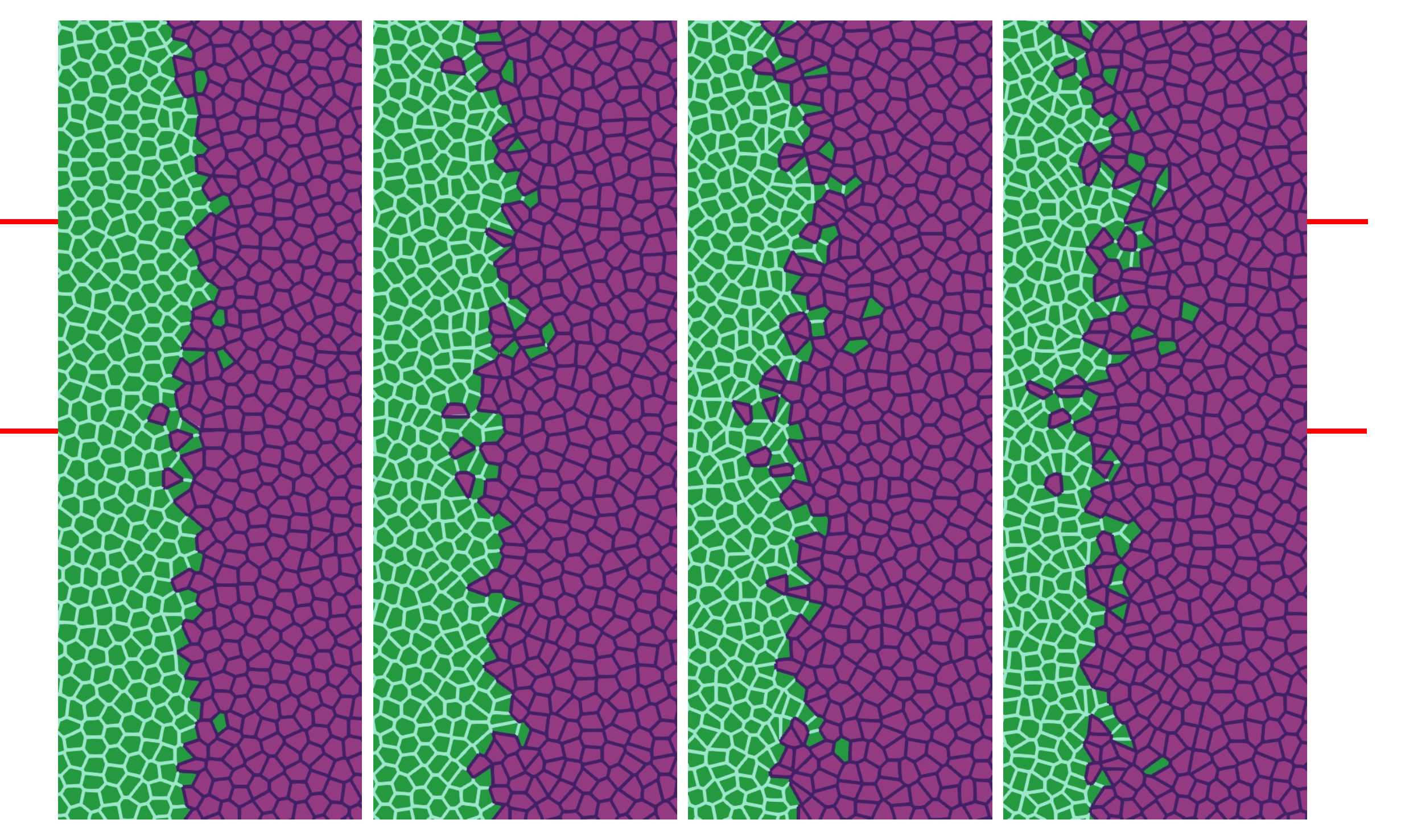}
\end{center}
\caption{Simulation of the antagonistic migration assay: one population advances pushing back the other, while the interfaces mix, creating scattered islands of cells of different type.  Parameters and times are as in Fig.~\ref{fig12}, except that one fifth of the overall population (randomly placed green and magenta cells) have $\Lambda_{12}=-7.5$ (population mixing) and the other four fifths have $\Lambda_{12}=-6.0$ (population segregation). The marked region has similar size to that reported in experiments \cite{moi19} and will be used in our TDA studies.
 \label{fig13} }
\end{figure}

The AVM allows us to study tissues that behave differently. In our simulations, 5000 cells are split into two populations with different properties specified by the junction tensions $\Lambda_{ij}$, $j=1,2$, which affect each pair of cell-cell contacts. The simulations producing Figs.~\ref{fig11}, \ref{fig12} and \ref{fig13} have open boundaries because we have focused on the interface between populations. We have fixed $K=\Gamma=1$ and $-6.8= \Lambda_{22} < \Lambda_{11} = -6.2$, which produce shape parameters $p^0$ of 3.50 (green cells) and 3.84 (magenta cells), below and above the transition value $p^{0*}=3.812$, respectively. Thus, Ras magenta cells are fluidlike (supercritical shape index) and their density is larger than that of the solidlike wt cells. This is consistent with the observation that wt cells have larger mean traction force amplitudes than Ras cells \cite{moi19}. Our aim is to analyze the effect of $\Lambda_{12}$ on the AMA.  Both monolayers occupy the right and left portions of a 4.4 mm wide, 3.1 mm tall box. In Figs.~\ref{fig11}-\ref{fig13}, we show a 1 mm $\times 2.5$ mm region.

In our simulations, we start from having the cell populations separated by a straight vertical interface at $L(0)=0$. The active force $\boldsymbol{\varphi}$ pushes Ras cells with $x>L(0)$ to the left, whereas $\boldsymbol{\varphi}=\mathbf{0}$ for any cell to the left of $x=L(0)$. The junction tension $\Lambda_{12}$ in Fig.~\ref{fig11} ($\Lambda_{12}=-7.0$) and in the two left panels of Fig.~\ref{fig12} ($\Lambda_{12}=-7.5$) favors population mixing. Ras (magenta) cells push wt (green) cells backwards at a velocity close to the observed $V^\text{interface}$, meanwhile creating a rugged interface between cell populations. As time elapses, fingers and some isolated islands (lagging wt in the Ras assembly and advancing Ras islands in the receding wt assembly) appear. These effects are more pronounced the smaller $\Lambda_{12}$ is, as shown by comparison of Figs.~\ref{fig11} and \ref{fig12}. It is possible to create some realistic mixing of the populations by changing the junction tension $\Lambda_{12}$ with time. The first two snapshots in Fig.~\ref{fig12} have $\Lambda_{12}=-7.5<\frac{1}{2}(\Lambda_{11}+\Lambda_{22})$, which favors population mixing. Then the interface between cell populations becomes very rugged and there appear islands of one cell type inside a layer of the other type. The third and fourth snapshots in Fig.~\ref{fig12} have been obtained with $\Lambda_{12}=-6.0> \frac{1}{2}(\Lambda_{11}+\Lambda_{22})$ that favors population segregation. The interface becomes smoother and the islands shrink and tend to disappear. 

We have also focused on the effects of cellular alignment. There are two terms in Eq.~\eqref{eq8} that try to synchronize cell velocities: the term proportional to $\beta$ and the active force $\boldsymbol{\varphi}$, which pushes the Ras cells to the left. Although the values of $\beta$ used to draw Figs.~\ref{fig11}-\ref{fig13} are smaller than that in Table~\ref{t1}, different $\beta$ still make a difference in the behavior during tissue collision, specially in the Ras population. Fig.~\ref{fig11} exhibits global polar migration because its $\beta$ value is larger than that in Figs.~\ref{fig12} and \ref{fig13}, but types of cells are not mixed despite having a favorable value $\Gamma_{12} = -7.0$. The smaller value of $\beta$ in Figs.~\ref{fig12} and \ref{fig13} creates a weaker polar alignment than that in Fig.~\ref{fig11}. The different patterns observed in these figures illustrate that cell alignment affects importantly the shape and configuration of the interface. Videos 3-5 in the supplemental material compare the time dynamics of these three sets of simulations.

\begin{figure}[h]
\begin{center}
\includegraphics[width=12cm,angle=0]{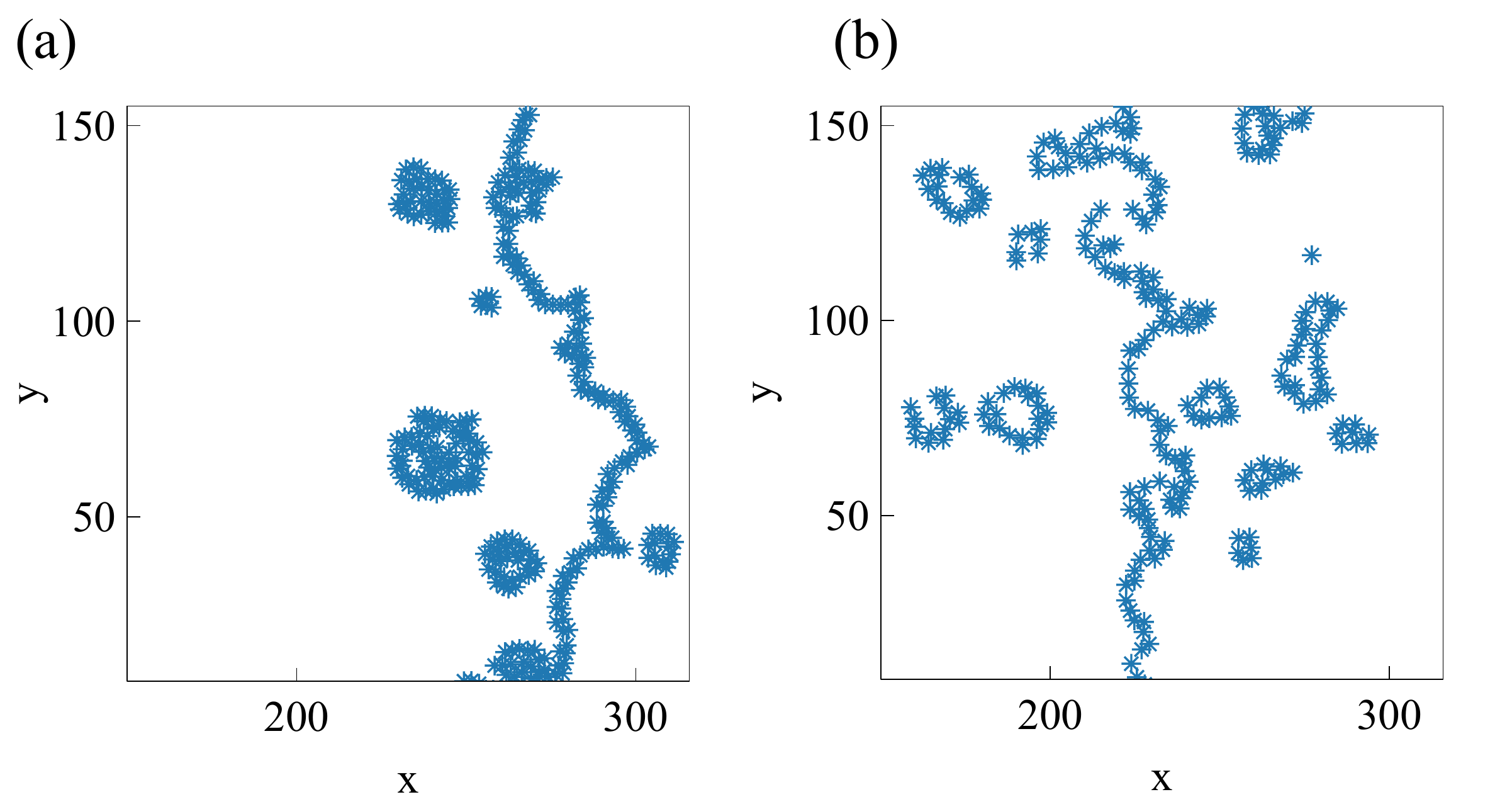} 
\end{center}
\caption{ Structure of the interface between colliding layers corresponding to two snapshot sof the collision of two confluent cellular monolayers in Moitrier {\em et al}'s experiment \cite{moi19}. These profiles correspond to (a) the third  and (b) the fourth panels (counting from the left) in the cover of Soft Matter corresponding to Ref.~\cite{moi19}.} \label{fig14}
\end{figure}

While the rightmost panel of Fig.~\ref{fig12} is similar to some of the experimental data \cite{moi19}, we can obtain a similar formation of islands and fingers by assuming that $\Lambda_{12}$ is randomly distributed among cells. In particular, we assume that one fifth of magenta and green cells have $\Lambda_{12}=-7.5$, which favors mixing of populations, while the remaining ones have $\Lambda_{12}=-6.0$ and favor population segregation. The result is depicted in Fig.~\ref{fig13}, which exhibits behavior similar to experimental observations \cite{moi19}, compare also to Figure \ref{fig14}. The topological data analyses of the next section characterize the geometry of the interface between cell types in antagonistic migration assays.

\section*{Formation of islands and Topological Data Analysis}\label{sec:4}
Experiments and numerical simulations  of cell monolayers produce time series of images 
that make it possible to identify the structure of interfaces and to compare their time evolution. It is quite cumbersome to process manually these time series. Here we use Topological Data Analysis (TDA) as a computational tool to process automatically time series of images. We next illustrate how to use TDA for this purpose and how to interpret the obtained results. We focus on specific parts of selected snapshots of images from experiments and then on time series of images from numerical simulations. While we have few images of interfaces from experiments, we can generate arbitrarily many from numerical simulations. Having many images, the automatic TDA tool enables us to describe in detail the topological changes of the interfaces and to implement hierarchical clustering strategies, thereby classifying the evolving interface structures.

Fig.~\ref{fig14} shows the interfaces between two colliding confluent cellular monolayers in an AMA \cite{moi19}. In this experiment, magenta Ras cells make green wild type cells move back, cf. third and fourth snapshots in the cover of Soft Matter, vol.\ {\bf 15} \cite{moi19}. The interface between the two cell populations is rather rough, it exhibits fingers, and there are islands or pockets of green cells left behind by the advance of the magenta front. To quantify these phenomena in an automatic way, we proceed as follows. Using Matlab, we  transform the images in matrices of ones (green) and zeros (magenta). Then we extract the positions of green/magenta interfaces, represented by the point clouds shown in Fig. \ref{fig14}, and process them using TDA. We pursue a similar strategy for images extracted from numerical simulations of our underdamped AVM, which yields a more complete picture of the evolution of interfaces.

\subsection*{Persistent homology}
A finite set of data points may be considered a sampling from the underlying topological space. Homology distinguishes topological spaces (e.g., annulus, sphere, torus, or more complicated surface or manifold) by quantifying their connected components, topological circles, trapped volumes, and so forth. Persistent homology characterizes the topological features of clouds of point data or particles at different spatial resolutions \cite{hat02}. Highly persistent features span a wide range of spatial scales. Persistent features are more likely to represent true features of the data/pattern under study than to constitute artifacts of sampling, noise, or parameter choice \cite{car09}. To find the persistent homology of a cloud of point data/set of particles, we must first view them as a simplicial complex $C$. Roughly speaking, a simplicial complex is defined by a set of vertices (points or particles) and collections of $k$-simplices. The latter are the convex hulls of subsets with $k+1$ vertices, comprising also faces; see the Appendix for precise definitions. Defining a distance function on the underlying space (the euclidean distance, for instance), we can generate a filtration of the simplicial complex, which is a nested sequence of increasingly bigger subsets. More precisely, a {\it filtration} of a simplicial complex $C$ is a family of subcomplexes $\{ C(r) \; \big| \; r \in \mathbb R \}$ of $C$ such that $C(r) \subset C(r')$ whenever $ r \leq r'.$ The filtration value of a simplex $S \in C$ is the smallest  $r$ such that $S \in C(r).$ The motivation for studying the homology of simplicial complexes is the observation that two shapes can be distinguished by comparing their holes. For $k \in \mathbb N$, the Betti number $\mathsf{b}_k$ counts the number of $k$-dimensional holes. A $k$-dimensional Betti interval $[r_b, r_d)$ represents a $k$-dimensional hole that is created at the filtration value $r_b$, exists for $r_b \leq r < r_d$ and disappears at value  $r_d.$ We are interested in Betti intervals that persist for a large filtration range: They describe how the homology of $C(r)$ changes with $r$.

\begin{figure}[h]
\begin{center}
\includegraphics[width=10cm]{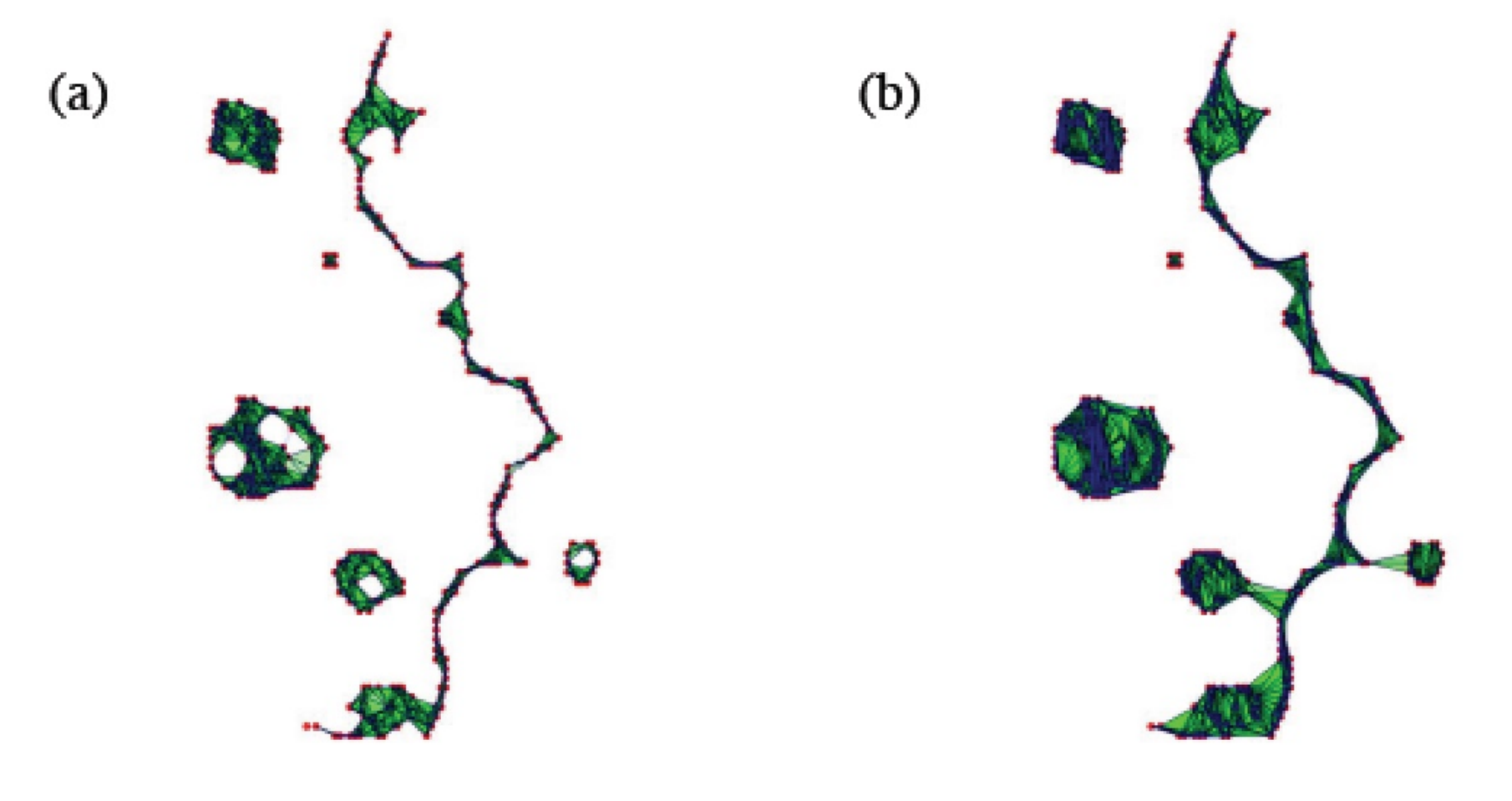}
\end{center}
\caption{Visualization of the complexes $VR(X,r)$ for the point cloud depicted in 
Fig.~\ref{fig14}(a)  when (a) $r=6$ and (b) $r=10$. For large enough $r$ all the components merge in a single one. Holes appear and disappear as new connections are created, reflecting the overall point cloud arrangement.
\label{fig15} }
\end{figure}

How do we construct a filtration? The Vietoris-Rips filtration $VR(X,r)$ \cite{hat02,car09}, 
which we will use here,  is constructed as follows:
\begin{itemize}
\item The set of vertices  $X$ is the cloud of points under study.
\item Given vertices $x_1$ and $x_2$, the edge $[x_1,x_2]$
is included in $VR(X,r)$ if the distance $d(x_1,x_2) \leq r$.
\item If all the edges of a higher dimensional simplex are included
in $VR(X,r)$, the simplex belongs to $VR(X,r)$.
\end{itemize}
A default choice for the distance $d$ to study homology of 2D particle 
configurations is the Euclidean metric.
Figure~\ref{fig15} displays two simplexes of a Vietoris-Rips filtration
for the point cloud in Figure~\ref{fig14}(a). Notice the appearance and
disappearance of holes and isolated components as the threshold 
distance $r$ to connect points increases.
This filtration is governed by three parameters:
\begin{itemize}
\item The maximum dimension $d_{max}$. This is the maximum dimension of the 
simplices to be constructed. The persistent homology (characterized by its Betti 
numbers) can be computed up to dimension $d_{max}-1.$ In this case $d_{max}=2$, 
we consider points ($0$-simplices), edges ($1$-simplices), and triangles 
($2$-simplices).
\item The maximum filtration value $r_{max}$ and the number of divisions $N$. 
These values define the filtered simplicial complexes to be constructed, for
$r \in \left\{
0, {r_{max} \over N-1}, {2r_{max} \over N-1},
\ldots, {(N-2) r_{max} \over N-1}, r_{max} 
\right\}.$
\end{itemize}
Notice that for a set of $P$ points, the full simplicial complex will have about $2^P-1$ simplices in it. Therefore, $d_{max}$ and $r_{max}$ are usually slowly increased to get information without reaching computational limits. The computation is not too sensitive to the specific value of $N$. When $r_{max}$ is greater than the diameter of the point cloud, all possible edges form and join all the points in one simplex.

For the readers' ease of use, we include more detailed definitions and intuitive examples in an Appendix. In the next two sections, we apply TDA to experimental and numerical images.

\subsection*{TDA of experiments.} 

\begin{figure}[!]
\begin{center}
\includegraphics[width=12cm]{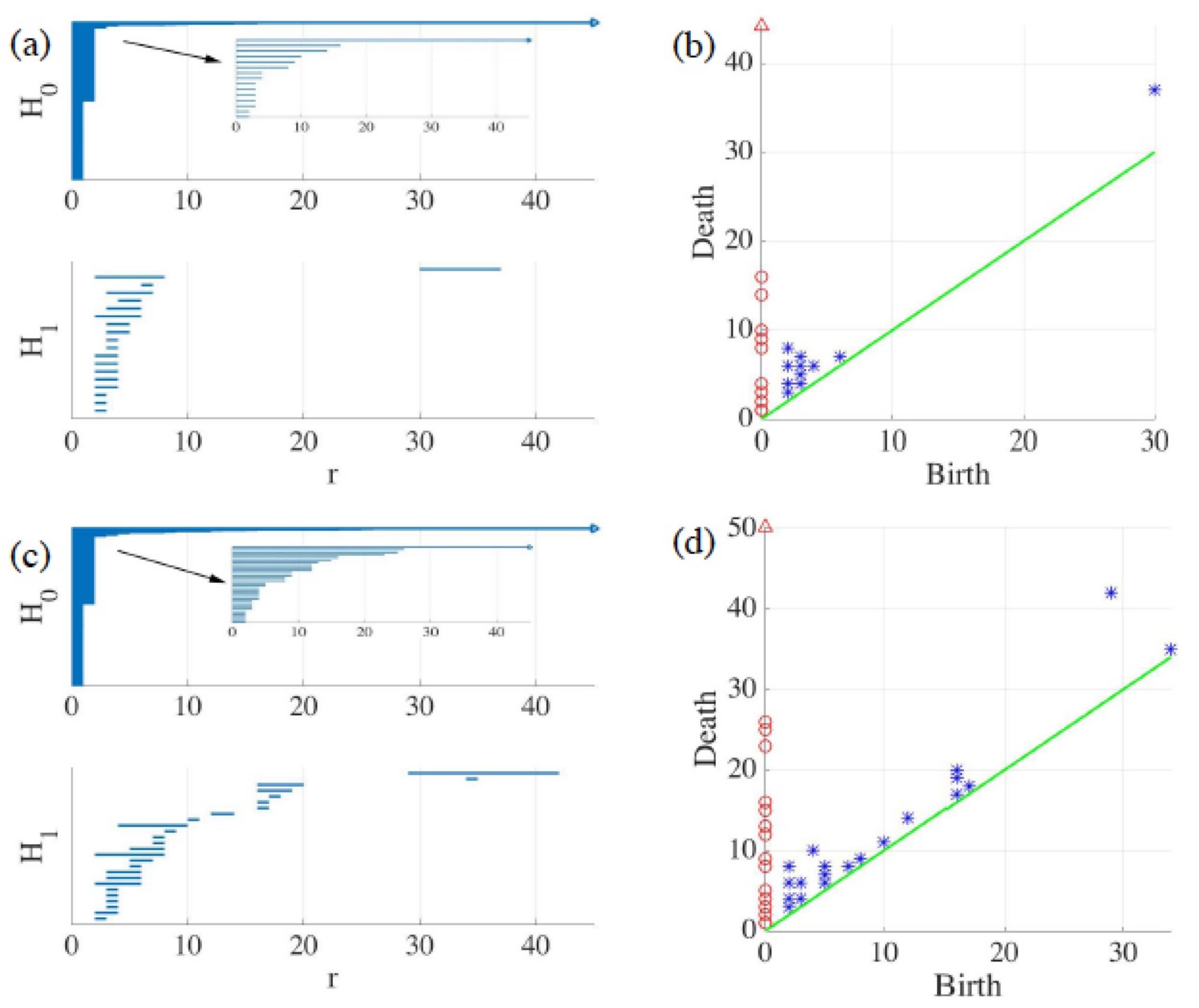}
\includegraphics[width=12cm]{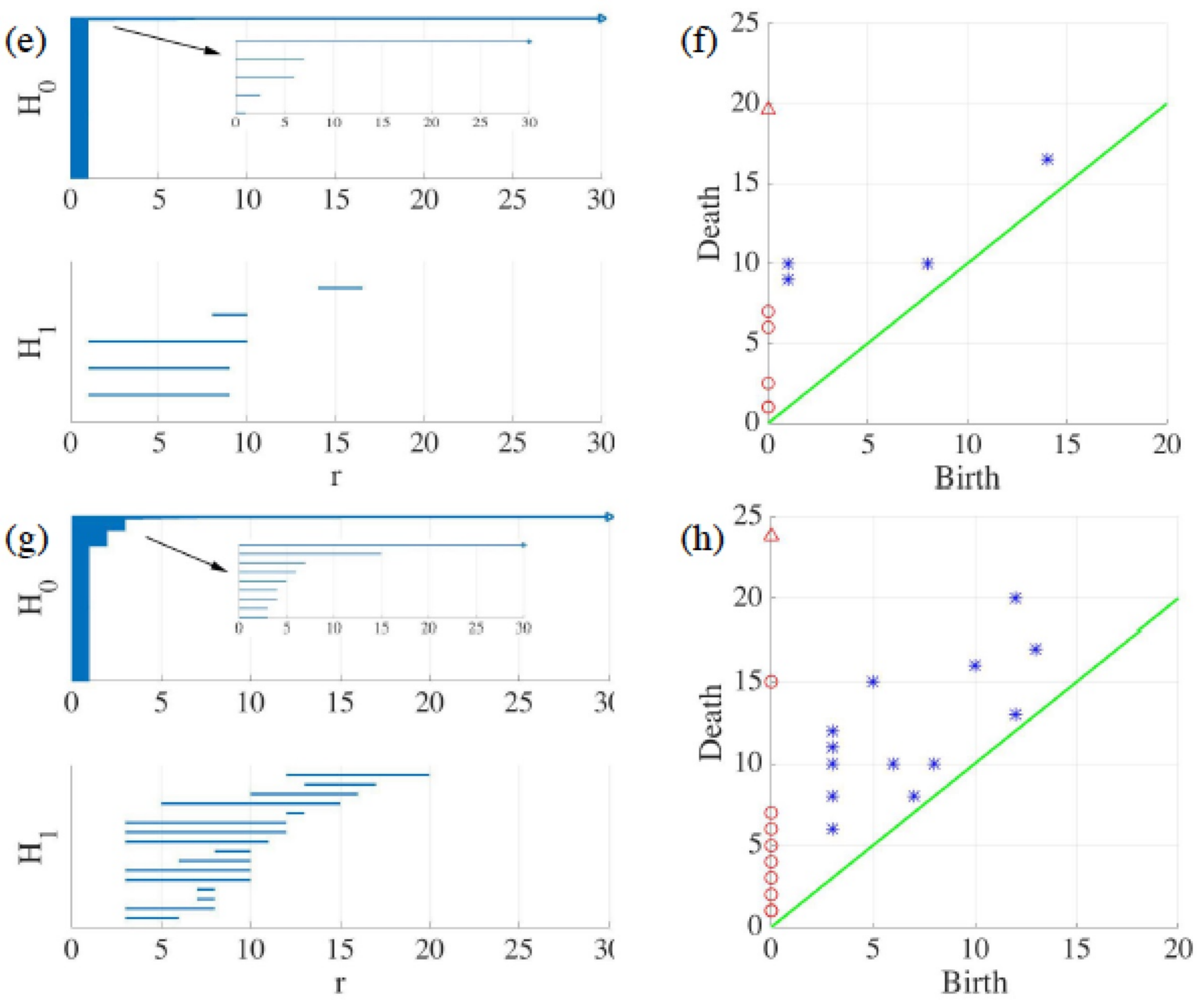}
\end{center}
\caption{Barcodes (left) and persistence diagrams (right) for the homologies $\mathsf{H}_0$ (circles) and  $\mathsf{H}_1$ (asterisks) of the interfaces separating cell types in images from experiments and numerical simulations. We use Vietoris-Rips filtrations with parameters $N$ and $r_{\rm max}$.  (a)-(b) TDA from Fig.~\ref{fig14}(a) (experiments) with $N=45,$ $r_{\rm max}=45$; (c)-(d) TDA from Fig.~\ref{fig14}(b)  (experiments) with $N=45,$ $r_{\rm max}=45$; (e)-(f) TDA from the leftmost panel in Fig.~\ref{fig13} (numerical simulations) with $N=60,$ $r_{\rm max}=30$; (g)-(h) TDA from the rightmost panel in Fig.~\ref{fig13} (numerical simulations) with $N=30,$ $r_{\rm max}=30$.  Points in the persistence diagrams mark the beginning (birth) and end (death) of a bar (homology class) in the barcode. Triangles represent
a component with infinite persistence. The green line is the diagonal.}
\label{fig16}
\end{figure}

\begin{figure}[h]
\begin{center}
\includegraphics[width=16cm]{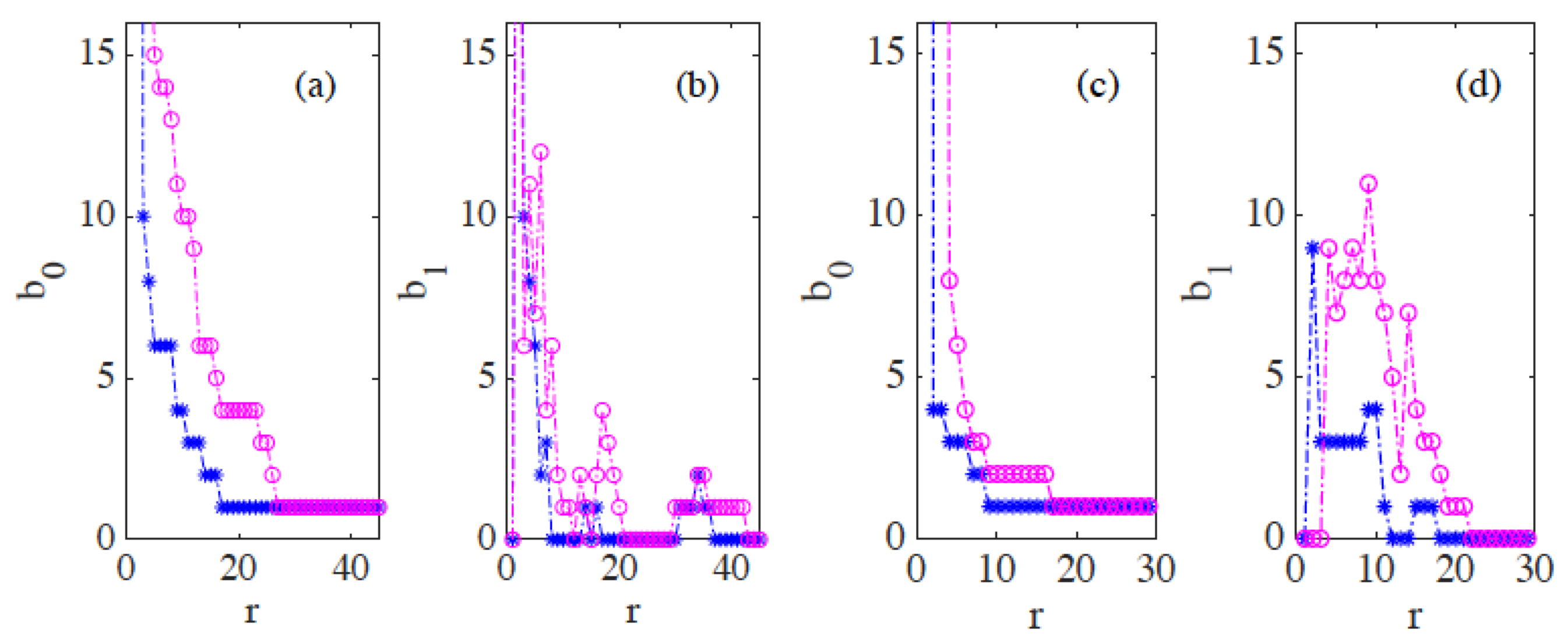}
\end{center}
\caption{(a)-(b) Betti numbers versus filtration parameter diagrams for Fig. \ref{fig14}(a) (blue asterisks) and \ref{fig14}(b) (magenta circles, later time in the AMA experiment) show that the number of clusters and holes in the interface between aggregates increases with time. (c)-(d) Same for the numerical simulations considered in Fig.~\ref{fig16}(e)-(h) corresponding to the leftmost and rightmost panels in Fig.~\ref{fig13}. As a result of island formation and motion, which increases with time, Panels (a) and (c) show that the number of components decreases more slowly with $r$ for the later time. The peaks in Panels (c) and (d) are similar for $r$ below $20$. In both cases, the interfaces formed at the later time display  larger numbers of holes with larger sizes as a result of island formation. The additional peaks in Panel (b) near $r=40$ correspond to islands that have already penetrated further inside the other cell population in the experiment. } 
\label{fig17}
\end{figure}

Let us consider the snapshots depicted in Fig. \ref{fig14}. Fig.~\ref{fig15} processes the earlier snapshot depicted in Fig.~\ref{fig14}(a), in which the green and magenta monolayers have made contact and started interpenetrating each other. Ras cells (magenta) are pushing back wt cells (green) towards the left. As they do so, there are islands of wt cells inside the Ras monolayer. How does TDA capture these features? After constructing the Vietoris-Rips filtration, there are two commonly employed graphical representations that visualize the persistent  homology of a point cloud: barcodes and persistence diagrams \cite{top15}.

Barcodes of a homology $\mathsf{H}_k$ depict Betti intervals $[r_b, r_d)$ for $k$-holes ($k>0$) or connected components ($k=0$) as the filtration parameter $r$ varies. The homology class $\mathsf{H}_0$ comprises the points forming the green/magenta interfaces. As the size filtration parameter $r$ increases from zero, there appear edges joining these points, thereby forming clusters as illustrated by Fig.~\ref{fig15} for specific values of $r$ and indicated by the barcodes in Fig.~\ref{fig16}(a) for the selected range of $r$. The class $\mathsf{H}_1$ further distinguishes compact components of the interface that are detached from the main part of the interface and form topological cycles, cf. the corresponding barcode in Fig.~\ref{fig16}(a). These components are islands of one cell type (phase) inside the bulk of the other phase. 

Persistence diagrams represent the Betti intervals by points in a birth-death plane (see Appendix for precise definitions). The $x$  axis represents the filtration value $r$ at which components/holes are created. The $y$ axis  represents the filtration value $r$ at which they disappear. Those points less close to the diagonal (green) tend to mark robust underlying geometrical features. Fig.~\ref{fig16}(b) depicts the persistence diagram corresponding to Fig.~\ref{fig14}(a). Red circles mark connected components of the interface between cell monolayers and the magnitude of the filtration parameter $r$ at which they disappear. As the filtration parameter increases, points comprising the main front merge rapidly in one component that absorbs neighboring clusters. They correspond to blocks of bars in the $\mathsf{H}_0$ panel of Fig.~\ref{fig16}(a) that start at the lowest value of $r$. Blue asterisks represent the appearance (horizontal axis) and disappearance (vertical axis) of holes inside such clusters. The first column of asterisks represents the ten bars in the $\mathsf{H}_1$ panel of Fig.~\ref{fig16}(a) that start at the same value of $r$ and form four groups of bars, which end at about the same value of $r$. The remaining bars and asterisks are similarly related. They represent the new holes that form as the clusters merge, which gives an idea of the relative arrangement thereof. Relatively narrow barcodes produce points in the persistence diagram that are close packed. 

Figs.~\ref{fig16}(c)-(d) display the barcodes and persistence diagram corresponding to Fig.\ref{fig14}(b). Compared to the earlier snapshot of Fig.~\ref{fig14}(a) and its TDA in Fig.~\ref{fig16}(a)-(b), there are more islands of each phase in the bulk of the other: the invasion of Ras cells leaves pockets of wt cells inside their midst. The main interface has become more meandering and exhibits more fingers than in the earlier snapshot. As a consequence, the number of clusters or interface components is larger than at the earlier time. Similarly, there are more topological cycles, which reflects the larger number of islands of one cell type in the midst of the other cell type. Barcodes and persistence diagram are more spread out. This is further quantified by the Betti numbers $\mathsf{b}_j$ that count the number of elements in $\mathsf{H}_j$, for $j=0$ (clusters) and for $j=1$ (holes), as depicted in Figs.~\ref{fig17}(a)-(b) for the snapshots shown in Fig. \ref{fig14}. The trends are similar in the simulations, as shown in Figs.~\ref{fig17}(c)-(d).

How do we characterize quantitatively variations in the persistence diagrams of point clouds? We have to introduce distances between diagrams to measure their differences. In the next section, we explain the bottleneck and Wasserstein distances using time series of numerical simulations, out from which we have generated much more complete data sets than available from experiments \cite{moi19}.

\subsection*{TDA of numerical simulations.} As indicated in the previous section, to observe island formation, we have to tune the (negative) junction tensions when simulating antagonistic migration assays. In particular, $\Lambda_{12}<\frac{1}{2}(\Lambda_{11}+\Lambda_{22})$ facilitates mixing of wt and modified cell populations whereas $\Lambda_{12}>\frac{1}{2}(\Lambda_{11}+\Lambda_{22})$ produces population segregation. In Fig.~\ref{fig12}, $\Lambda_{12}$ switches from population mixing to segregation after the two first snapshots. Then the pockets of green cells left behind by the advance of the interface shrink and start disappearing, as shown in the third and fourth snapshots of Fig.~\ref{fig12}. If mixing is weaker, as in Fig.~\ref{fig11}, the interface forms pronounced fingers, there are less islands and we do not need to change the junction tensions with time. In Fig.~\ref{fig13}, $\Lambda_{12}$ randomly takes on a mixing value for one fifth of Ras and wt cells and on a segregation value for the others. The results of changing interface and island formation are qualitatively similar to those observed in experiments. 

Let us now interpret the evolution shown in the panels of Fig.~\ref{fig13} using TDA (see also Video 5 in the supplemental material, out from which we have extracted 12 snapshots). Figures \ref{fig16}(e)-(h) and \ref{fig17}(c)-(d) show the barcodes, persistence diagrams and Betti numbers for the marked sections of the leftmost and rightmost panels in Fig.~\ref{fig13}. As before, we represent the interfaces by point clouds (datasets are included in the supplemental material). At $r=0$, each point of the interface is a component. For the more regular interface of the leftmost panel in Figure \ref{fig13}, increasing $r$ produces point components appearing as the short $\mathsf{H}_0$ bars in Figure \ref{fig16}(e). These bars end at similar filtration values and appear as a single red circle in the persistence diagram of Figure \ref{fig16}(f). The main three islands correspond to the three intermediate bars in the inset of Fig.~\ref{fig16}(e), which disappear at larger filtration values. The lowest circle in Fig.~\ref{fig16}(f) represents the point components, the three intermediate ones represent the islands in the barcode and their sizes. All clusters finally merge in the main front represented by the arrow on top of the vertical axis in Fig.~\ref{fig16}(f). Analysis of $\mathsf{H}_1$ confirms that the intermediate circles/bars are round islands and not strings. Each component corresponds to a cycle represented by the three largest $\mathsf{H}_1$ bars in Figure \ref{fig16}(e) and the two first asterisks in Figure \ref{fig16}(f), one of which represents the two bars of similar length. The two shortest bars represent holes formed as components merge during the filtration process and  correspond to the two asterisks closer to the diagonal in Figure \ref{fig16}(f). 

Figs.~\ref{fig16}(g)-(h) correspond to the more meandering interface of the rightmost panel in Fig.~\ref{fig13}. There are more points in the cloud representing the interface, whose irregularity results in different extinction values of $r$ for the associate  $\mathsf{H}_0$ bars. The main seven islands correspond to the intermediate bars in the inset of Fig.~\ref{fig16}(g), and their extinction values in the persistence diagram give an idea of the distance to the main front or to another island. The fact that they are islands (enclosed by a boundary) is inferred from the $\mathsf{H}_1$ bars in Fig.~\ref{fig16}(g). They correspond to the seven bars that appeared first, which are also represented by the first column five asterisks in Fig.~\ref{fig16}(h) having smaller $r$. Two of the asterisks correspond to two islands of similar size length each, which have bars of similar size. The length of the bars in the barcode or the distance of the asterisks from the diagonal in the persistence diagram give an idea of the island size. Additional $\mathsf{H}_1$ bars represent holes created during the filtration process as components merge and give an idea of the relative arrangement of the islands or of the fingers in the main front. They are represented by the additional asterisks in Fig.~\ref{fig16}(h). The Betti numbers in Fig.~\ref{fig17}(c)-(d)  show a larger number of island and holes as time increases from the leftmost snapshot in Fig.~\ref{fig13} to the rightmost one. Compared to the TDA of experiments in Figs.~\ref{fig16}(a)-(d), there are no gaps between bars and asterisks appearing for large $r$ in Fig.~\ref{fig16}(e)-(h). The reason is that the distance of islands to the main front is smaller for the simulation than for the experiment.  

\begin{figure}[h!]
\begin{center}
\includegraphics[width=13cm]{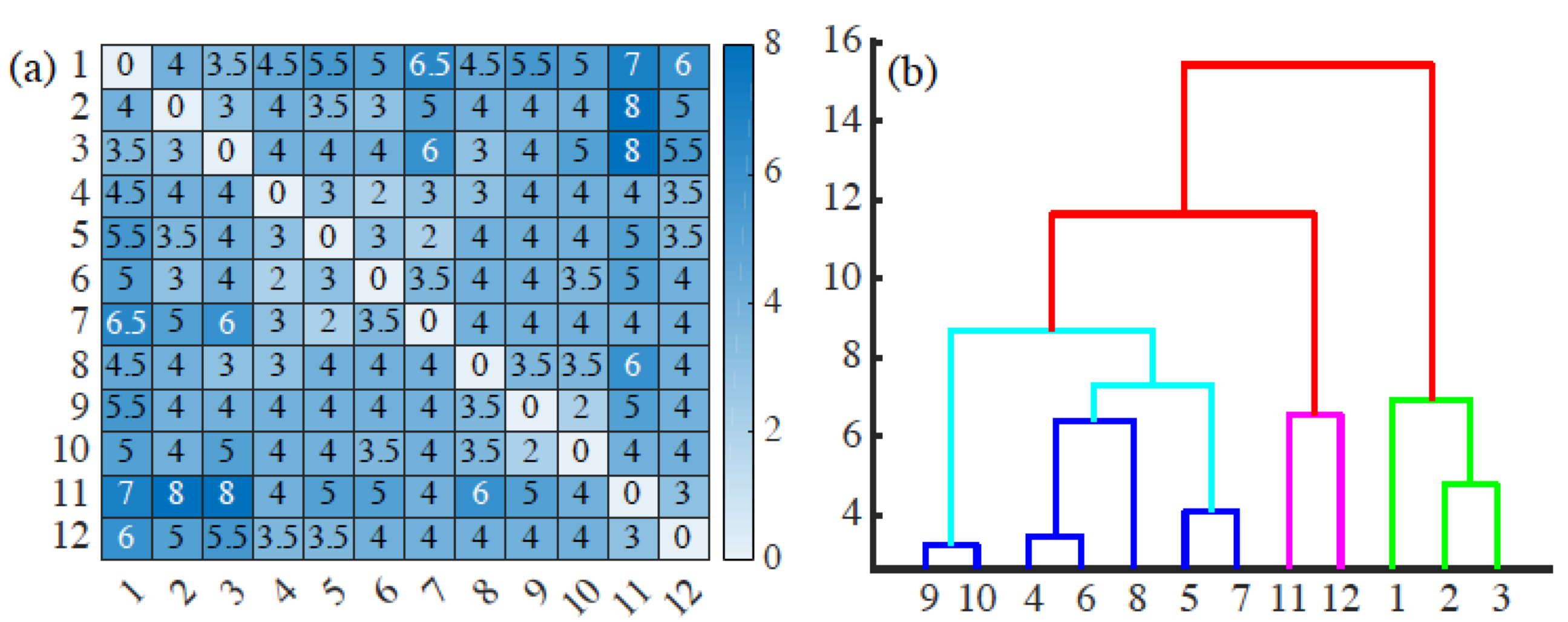}
\end{center}
\caption{(a) Bottleneck distance matrix for the interfaces between cells populations appearing in the $12$ snapshots forming Video 5 in the supplemental material and (b) associated dendrogram illustrating how the interfaces between cell populations can be grouped in clusters. Interfaces in frames $1$-$3$, $4$-$10$, $11$-$12$ can be grouped together, and the last two groups are closer to each other than to the initial frames. These groupings reflect similarities between frames as they succeed one another, and the disruptions between frames reflect significant topological changes of the interfaces (e.g., detachment and reattachment of islands).}
\label{fig18}
\end{figure}

We have applied TDA to a time series of 12 snapshots (extracted from Video 5 in the supplemental material), which visualize the evolution of the numerically simulated interface in an AMA. Figures \ref{fig16}(e)-(h) correspond to snapshots $2$ and $10$. For each of them, we calculate barcodes as explained above. To quantify the variations in the barcode patterns, we introduce distances between persistence diagrams that are stable against random perturbations \cite{ker17,ede10}. Given two persistence diagrams $X$ and $Y$, their {\em bottleneck distance} is defined as
\begin{eqnarray}
W_\infty(X,Y)= \inf_{\varphi:X\to Y} \sup_{x\in X} \|x-\varphi(x)\|_\infty.  \label{eq13}
\end{eqnarray}
Here $\varphi$ ranges over all bijections between the persistence diagrams (taking the diagonal into account, see Appendix) and $\|x\|_\infty =$ max$_i\{|x_i|\}$ is the usual $L_\infty$-norm over the points $x$ of the persistent diagram $X$. The bottleneck distance is an example of the more general {\em Wasserstein distance} between persistent diagrams:
\begin{eqnarray}
W_{q,p}(X,Y)= \biggl[\inf_{\varphi:X\to Y} \sum_{x\in X} \|x-\varphi(x)\|_p^q\biggr]^{1/q}, \quad W_q(X,Y)=W_{q,\infty}(X,Y).  \label{eq14}
\end{eqnarray}  
We have $W_{\infty}(X,Y)=W_{\infty,\infty}(X,Y)$. Figure \ref{fig18}(a) represents the matrix of bottleneck distances between the persistence diagrams of the $12$ frames in Video 5. These distances are stable in the sense that a small perturbation in the input filtration leads to a small  perturbation of its persistence diagram in the bottleneck distances \cite{ede10} ($q$-Wasserstein distances share that property too). Efficient algorithms to compute these distances are discussed in \cite{ker17}. 
Techniques enabling us to input topological features into deep neural networks and learn task-optimal representations during training are proposed in \cite{hof17}. We could use a neural network approach if we wanted to relate them to a specific pattern, but that is not the case here. 

Once we have a matrix of distances, we resort to unsupervised clustering methods to classify the frames in similar blocks. This classification automatically extracts the similitudes and changes between interfaces at different times of the AMA. Figure \ref{fig18}(b) displays a dendrogram obtained by agglomerative hierarchical clustering using Ward's method \cite{agnes} and the bottleneck distance. A dendrogram consists of U-shaped lines that connect data points (which, in this case, are the interfaces in each frame) in a hierarchical tree. The height of each U represents the distance between the interfaces that are connected by it. A dendrogram is not a single set of clusters, but  a multilevel hierarchy. For a given dendrogram, we identify the natural cluster divisions relying on the inconsistency coefficient  \cite{inconsistency}. The latter compares the height of a link in a cluster hierarchy with the average height of links below it: larger inconsistency coefficients mark natural divisions  \cite{inconsistency}. By defining a cutoff value for the inconsistency coefficient, we automatically detect clusters.
A $0.9$ cutoff in the inconsistency coefficient detects $3$ clusters, corresponding to times $\{1,2,3\}$, $\{4,5,6,7,8,9,10\}$, and $\{11,12\}$ in Video 5. Thus, we  distinguish the initial period (interfaces close to the original connected one, times $\{1,2,3\}$), the intermediate period (a phase in which a few islands form and advance, times $\{4,5,6,7,8,9,10\}$), and the final period (severe disruption with the abrupt formation of several islands, times  $\{11,12\}$). In this way, we gain insight on the time evolution: how fast the interfaces experience significative changes, and when abrupt changes do occur and mark the onset of a new cluster of frames. The chosen cutoff 0.9 is a critical value: Increasing it, we find only one cluster. Lowering the cutoff, we detect $5$ clusters of frames, corresponding to times $\{1,2,3\}$, $\{4,6,8\}$,  $\{5,7\}$, $\{9,10\}$, $\{11,12\}$. With respect to the 0.9 value, the intermediate cluster splits into other $3$, reflecting detachment, reattachment  and slow progression of islands.

We can also obtain clusters by setting distance cutoffs, i.e., a height in the dendrogram of Fig.~\ref{fig18}(b). The height of a link between clusters represents the distance between them, which is called cophenetic distance. It is possible to calculate the correlation between the cophenetic distance and the distances in the matrix of Fig.~\ref{fig18}(a) (the cophenetic correlation \cite{cophenetic}). For our simulations, Ward's hierarchical clustering provides the largest cophenetic correlation when  compared with other clustering approaches, such as single-linkage \cite{mcg20}. Thus, the Ward dendrogram gives the most faithful representation of the distance matrix, which is why we use it. Depending on the chosen cutoff height in Fig.~\ref{fig18}(b), we obtain one, two, three, four or five clusters. The sets of three or five clusters are the same as before (this does not necessarily occur in general). Alternatively, we can use the K-means algorithm  to group the interfaces in clusters, selecting an optimal number of clusters by silhouette or  elbow type criteria \cite{kmeans,silhouette}. In our case, K-means with 3 or 5 clusters produces the ones already obtained (this does not necessarily occur in general). We have illustrated TDA with a relatively short time series, but it is clear that it could be used for automatic detection of topological changes in much longer time series, or to quantify how close the interfaces obtained from different simulations or experiments are. In the Appendix, we have used TDA with 1-Wasserstein distance to interpret the evolution of tissue interfaces in the numerically simulated spreading assay of Fig.~\ref{fig4}.

\section*{Conclusions}\label{sec:5} 
We have modeled how epithelial cell aggregates advance through empty spaces (wound healing, tissue spreading) and collisions between aggregates (tumoral invasion) using an active vertex model with dynamics for cell centers that includes collective tissue forces \cite{bar17}, and velocity alignment and inertia \cite{sep13}. The active vertex model implements exchanges of neighboring cells automatically (T1 transitions) and uses the SAMoS software. Compared with particle models with underdamped dynamics, our model accounts for fingering instabilities in spreading tissue without having to introduce leader cells \cite{sep13}. Compared to continuum models \cite{ale19}, stochasticity enables our model to reproduce the observed fast irregular oscillation of cell velocities in fingers \cite{pet11} and the spatial autocorrelation of the velocity \cite{ang10}. Our underdamped AVM predicts that cells at the interface and the fingers have larger area than those well inside the tissue, which has been corroborated by recent experiments \cite{lv20}. We also observe in numerical simulations of tissue spreading that the velocity of the fastest cell in a finger may oscillate with a short period in a range between 30 minutes to about one hour. A similar short period oscillation has been observed in experiments; cf Figure 101A in L. Petitjean's PhD thesis \cite{pet11}. Thus, for spreading tissue, detailed comparison to experimental data provides a quantitatively accurate description of cell motion (speed, velocity correlation function and polar order parameter). For antagonistic migration assays, we have reproduced collisions in which one cell population pushes back another whereas both populations mix forming different types of interfaces. The key element to model mixing is to keep different junction parameters for the two colliding tissues: the invading cells are liquid like whereas the receding tissue comprises solid like cells. In addition, a fraction of cells favor mixing, the others segregation, and that these cells are randomly distributed in space. Thus characterized, numerical simulations produce outcomes similar to those observed in experiments \cite{moi19}. Compared to particle models, ours includes active vertex forces between cells that keep them together preventing gaps and keeping track of cellular compression, enlargement and changes of area. To characterize automatically the dynamics of islands and the rugged interface between aggregates, we have introduced topological data analyses of experiments and time series from numerical simulations. In collisions between aggregates, the interface between wt and Ras cell populations roughens and islands appear. The persistence diagrams of Homology classes 0 (clusters) and 1 (cycles) spread out and the number of these classes given by the corresponding Betti numbers increases. Using time series of data generated by numerical simulations, we have explained how to cluster interfaces using distance matrices based on the bottleneck distance between their persistence diagrams, which are stable to perturbations in the process. Despite the amount of data from experiments being limited, disruptive events such as island and cluster formation can be automatically captured by topological data analyses of numerical simulations and contrasted with experiments. Similarly, the Wasserstein distance between images enables us to track and classify automatically the evolving shapes of interfaces between cell populations by using time series from experimental or numerical studies. These techniques of topological data analysis are scalable and could be used in studies involving large amounts of data whenever available. 

Our results allow to extract parameter values and to determine {\em biologically relevant} physical mechanisms for characterizing confluent motion of cellular aggregates, as described above. In particular: (i) cells at the interface are larger,  inform the aggregate motion and are influenced by it, without needing leader cells to form fingers at the interface; and (ii) in colliding cellular aggregates, the solid or liquid like character of the cells (as determined by their junction parameters) decides the way the invasion goes. These aspects of our model are important in ascertaining how the biophysical features of materials influence tissue/organ regeneration \cite{li17}. Our work provides researchers in the field with useful tools
to gain biological insight, to devise and to interpret data from experiments. To enhance the value of our results, e.g, for studies of metastatic cancer, future works may add cellular mechanisms such as Notch signaling dynamics \cite{boa15}, models of epithelial/mesenchymal transition and cancer stem cell formation \cite{boc17,boc18} to our vertex model. This venue has been successfully followed in studies of angiogenesis \cite{veg20}.


\section*{Acknoledgements}
Parts of this work were carried out at the Workshop on Modeling Biological Phenomena from Nano to Macro Scales, held in 2018 at the Fields Institute in Toronto, Canada. We thank Prof. I. Hambleton, director of the Fields Institute, for the invitation and support during the Workshop. We thank Rastko Sknepnek for his help in implementing SAMoS software and for related and useful comments during the Workshop.  The authors thank Russel Caflisch for hospitality during their stay at the Courant Institute of Mathematical Sciences, New York University, where the bulk of the work was carried out. This work has been supported by the FEDER/Ministerio de Ciencia, Innovaci\'on y Universidades -- Agencia Estatal de Investigaci\'on grants MTM2017-84446-C2-2-R (LLB and CT) and MTM2017-84446-C2-1-R (AC). CT thanks the Programa Propio of Universidad Carlos III de Madrid for a scholarship to finance her stay at the Courant Institute. 
 
\appendix
\section*{Topological data analysis}\label{ap1}
For the reader's ease of use, this Appendix makes more precise some definitions and includes simple examples to provide an intuitive idea of the meaning of persistent homology features. It is structured as follows. First, we give more precise definitions of persistent homology concepts. Second, we present applications to simpler synthetic data, for an easier visual interpretation of the results in the main text when interfaces are formed by many connected components. Finally, we discuss how to extract information on front roughness from numerical or experimental data, when interfaces define a single component.

\paragraph{Basic definitions of persistent homology and examples.}
As said before, a finite set of data points may be considered a sampling from the underlying  topological space. Data structure can be investigated by creating connections between proximate data points, varying the scale over which these connections are made, and looking for features that persist across scales \cite{top15}. Homology distinguishes topological spaces (e.g., annulus, sphere, torus, or more complicated surface or manifold) by quantifying their connected components, topological circles, trapped volumes, and so forth. Persistent homology describes how the homology of a nested family of simplicial complexes changes with respect to a defining parameter. What is  a simplicial complex $S$? To define it, we need three elements \cite{ghr14}:
\begin{itemize}
\item A set  of points $X$ in a space of dimension $D$.
\item Sets of $k$-simplices $[\nu_0, \nu_1, \ldots, \nu_k ]$ with vertices $\nu_i \in S$, $i=1, \ldots, k$, for each $k \geq 1$. A $k$-simplex is a $k$-dimensional polytope which is the convex hull of its $k + 1$ vertices:
\begin{eqnarray*}
[\nu_0,\nu_1,\ldots,\nu_k] = \left\{
\theta_0 x_0 + \theta_1 x_1 + \ldots +
\theta_k x_k \; \big| \; \sum_{i=1}^k \theta_i = 1,
\; \theta_i \geq 0, \; i=1,\ldots, k.
\right\}
\end{eqnarray*} 
The vertices must be affinely independent, i.e., the difference vectors $\nu_1-\nu_0,\ldots \nu_k-\nu_0$ must be linearly independent. The $k$-simplex is oriented so that an odd permutation of the points in $[\nu_0,\ldots,\nu_k]$ reverses its sign. 
\item A $k$-simplex has $k+1$ faces, each constructed by deleting one of the vertices. The faces must satisfy the following property: If $[\nu_0,\nu_1,\ldots,\nu_k]$ belongs to the simplicial complex $S$, then all its faces must also be in the simplicial complex $S$. This can be made more precise. The set of all $k$-simplices in $S$ is a vector space $\mathsf{C}_k$. The boundary of a $k$-simplex is the union of all its $(k-1)$-subsimplices. For each $k\geq 1$, the boundary map $\partial_k: \mathsf{C}_k\to \mathsf{C}_{k-1}$ is the linear transformation defined by 
\begin{eqnarray}
\partial_k([\nu_0,\ldots,\nu_k])=\sum_{j=0}^k (-1)^j [\nu_0,\ldots,\hat{\nu}_j,\ldots,\nu_k],    \label{a1}
\end{eqnarray}
where $[\nu_0,\ldots,\hat{\nu}_j,\ldots,\nu_k]$ is the $(k-1)$-simplex obtained by removing the vertex $\hat{\nu}_j$ from $[\nu_0,\ldots,\nu_k]$. 
\end{itemize}
The motivation for studying the homology of simplicial complexes is the observation that two shapes can be distinguished by comparing their topological features. A disk is not a circle because the disk is solid, while the circle has a hole. Similarly, a circle is not a sphere, because the sphere encloses a two dimensional hole, whereas the circle encloses a one dimensional hole. To distinguish topological features, we need several definitions. Boundary operators connect the vector spaces $\mathsf{C}_k$ into a {\em chain complex} $\ldots\to \mathsf{C}_{k+1}\to \mathsf{C}_k\to\mathsf{C}_{k-1}\to\ldots\to\mathsf{C}_0\to 0$. The kernel and image of boundary operators determine $k$-cycles $\mathsf{Z}_k=$Ker$\{\partial_k: \mathsf{C}_k\to\mathsf{C}_{k-1}\}$ and $k$-boundaries $\mathsf{B}_k=$Im$\{\partial_{k+1}: \mathsf{C}_{k+1}\to\mathsf{C}_k$, respectively. Since a boundary has no boundary \cite{ghr14}, $\mathsf{B}_k$ is a subspace of $\mathsf{Z}_k$. Thus, $\mathsf{C}_k$ is the vector space of all $k$-chains in the simplicial complex $S_r$, $\mathsf{Z}_k$ is the subspace of $\mathsf{C}_k$ consisting of $k$-chains that are also $k$-cycles, and $\mathsf{B}_k$ is the subspace of $\mathsf{Z}_k$ consisting of $k$-cycles that are also $k$-boundaries. We say that two $k$-cycles are homologous (equivalent) if they differ by a $k$-boundary. This equivalence relation splits $\mathsf{Z}_k$ in equivalence classes denoted by $[z]$ if $z\in\mathsf{Z}_k$. The $k$th homology of $S_r$ is the quotient set $\mathsf{H}_k=\mathsf{Z}_k/\mathsf{B}_k$ comprising all  equivalent $k$-cycles. The dimension of $\mathsf{H}_k$, $\mathsf{b}_k=$ dim$\mathsf{Z}_k-$dim$\mathsf{B}_k$, is the $k$th {\em Betti number}.  
In terms of the topological characteristics, $\mathsf{b}_k$ is the number of independent holes of dimension $k$. For instance, $\mathsf{b}_0$ is the number of connected components, $\mathsf{b}_1$ is the number of topological circles, $\mathsf{b}_2$ is the number of trapped volumes, and so on. The topology of a simplicial complex may be described by the sequence of Betti numbers, $\mathbf{b}=(\mathsf{b}_0,\mathsf{b}_1,\ldots)$. For instance, a topological circle has $\mathbf{b}=(1,1,0,\ldots)$, a topological torus has $\mathbf{b}=(1, 2, 1, 0,\ldots)$, and a topological sphere has $\mathbf{b}=(1,0,1,0,\ldots,)$. Betti numbers are a topological invariant, meaning that topologically equivalent spaces have the same Betti number.

\begin{figure}[h!]
\begin{center}
\includegraphics[width=13cm]{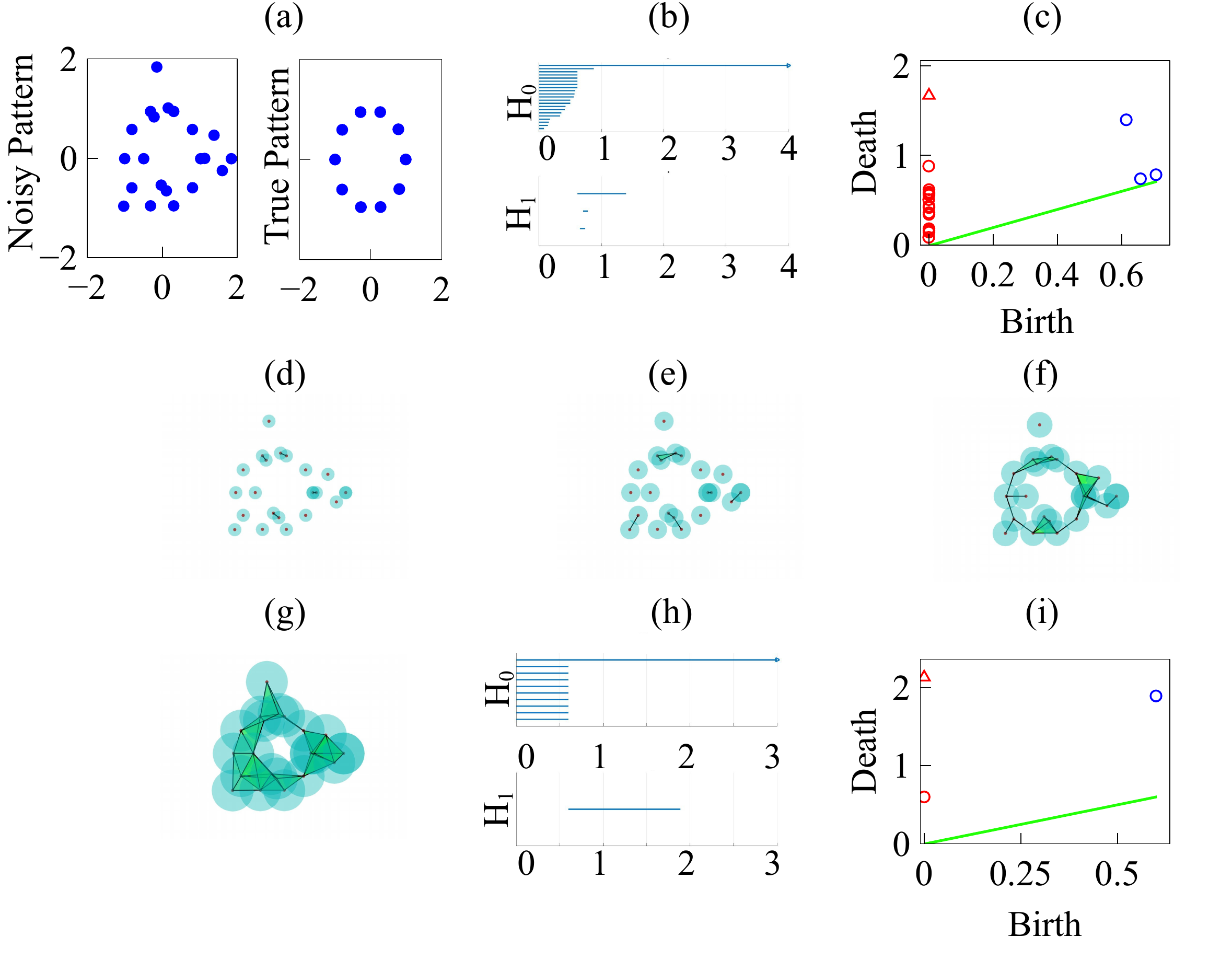}
\end{center}
\caption{Persistent homology for data on a circle. (a) Noisy versus true circle data. (b) Barcodes for the Betti numbers $\mathsf{b}_0$ ($\mathsf{H}_0$) and $\mathsf{b}_1$ ($\mathsf{H}_1$) for the noisy data. (c) Persistence diagram for the noisy data with $r_{max}=4$ and $N=100.$ (d)-(g) Vietoris-Rips simplicial complexes formed from the noisy data increasing the filtration parameter $r$. (h) Barcodes for the Betti numbers $\mathsf{b}_0$ and $\mathsf{b}_1$ for clean data on the circle. (i) Persistence diagram for clean data on the circle, with $r_{max}=3$ and $N=100.$} \label{fig19}
\end{figure}

\begin{figure}[h!]
\begin{center}
\includegraphics[width=13cm]{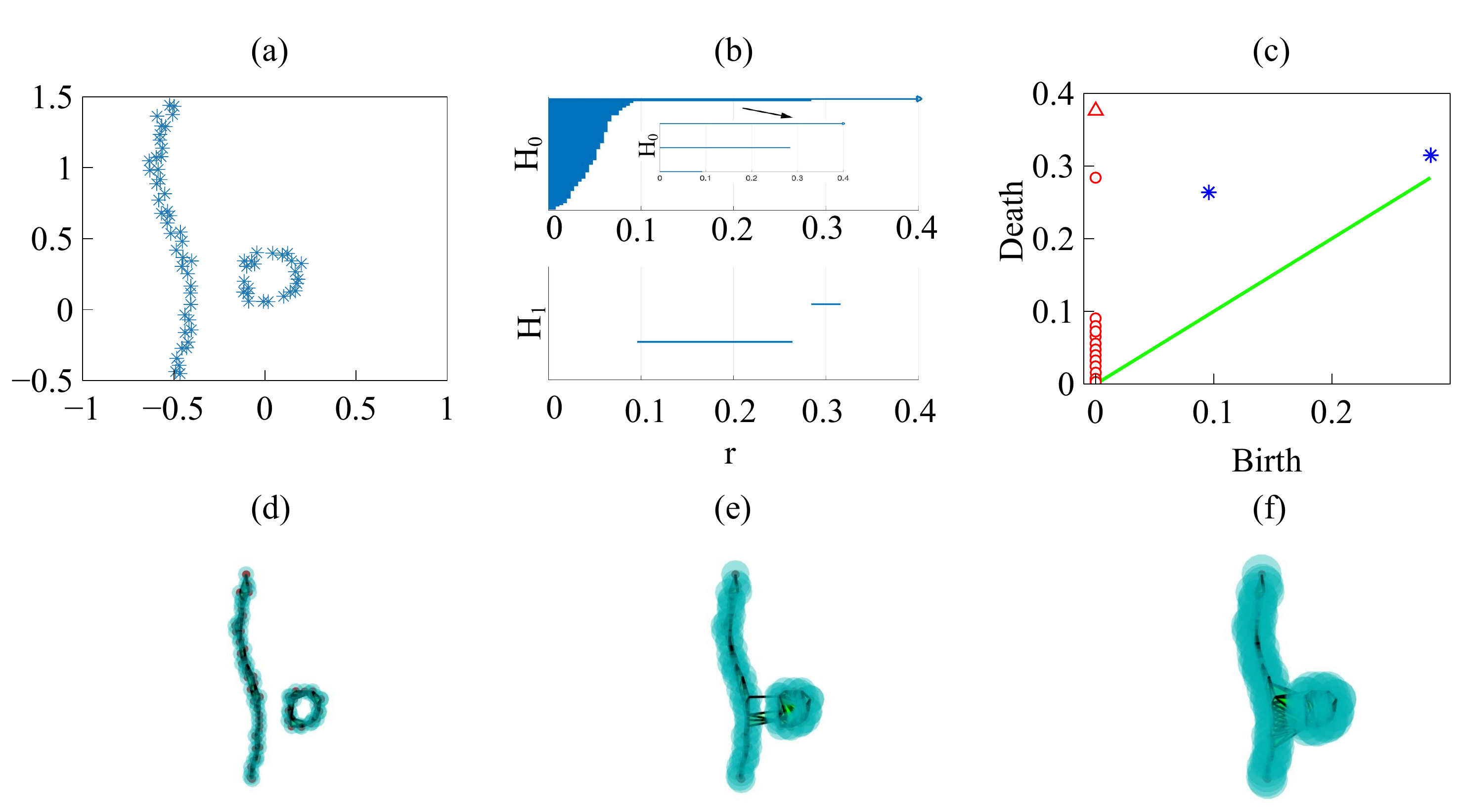} 
\end{center}
\caption{Persistent homology for the border of two colliding populations. We should stress that these examples use schematic figures, with clear fronts, not results from experiments or simulations (which have larger noise, and less clear features). Thus,  the persistent homology of schematic figures is clearer and easier to interpret. (a) Interface separating the two populations. 
(b) Barcodes for the Betti numbers $\mathsf{b}_0$ ($\mathsf{H}_0$) and $\mathsf{b}_1$ ($\mathsf{H}_1$), and (c) Persistence diagram with $r_{max}=0.4$ and $N=100$. (d)-(f) Vietoris-Rips simplicial complexes formed increasing the filtration parameter $r$.}
\label{fig20}
\end{figure}

\begin{figure}[h!]
\begin{center}
\includegraphics[width=13cm]{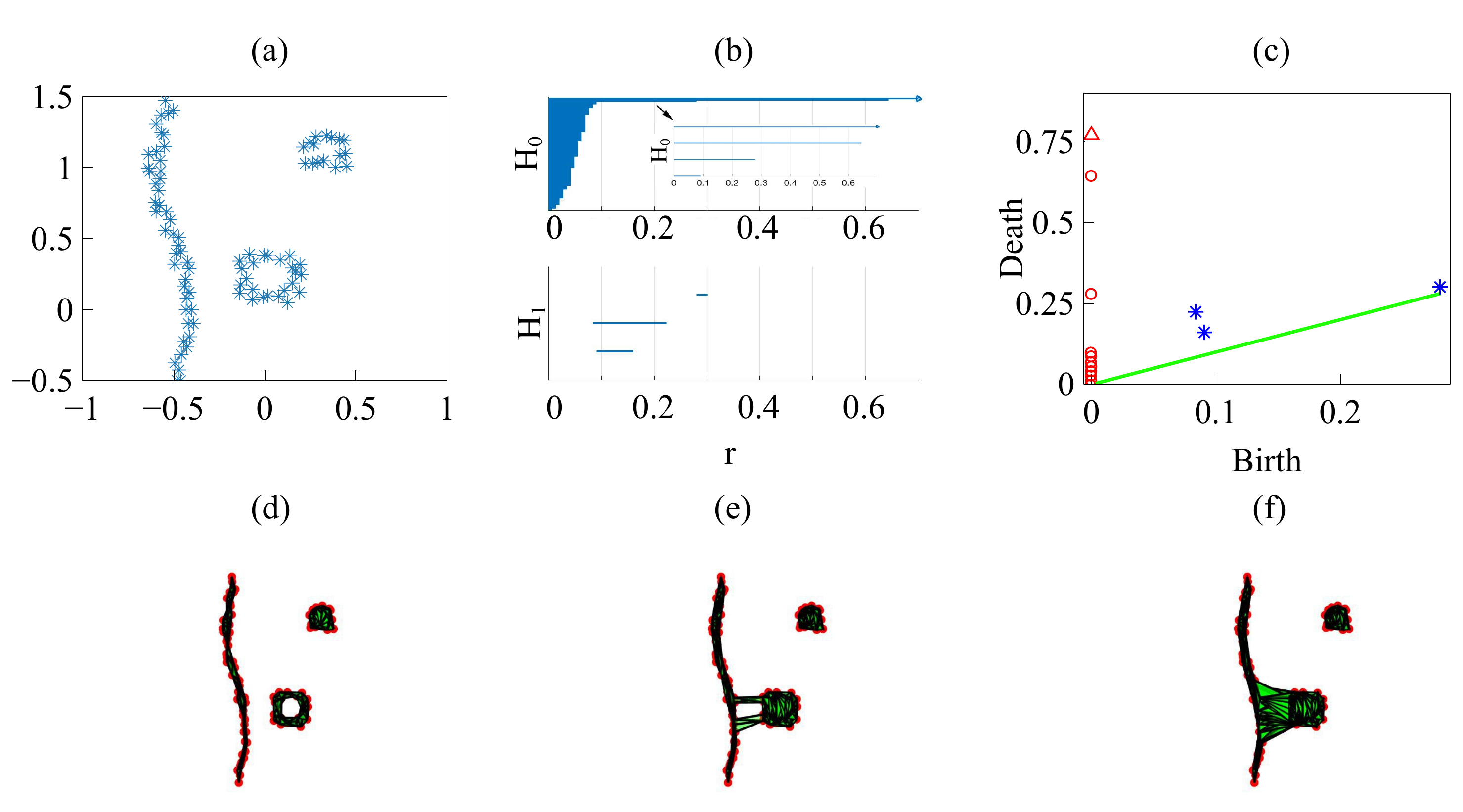} 
\end{center}
\caption{Same as in Fig.~\ref{fig20} except that there are one interface and two islands. 
Parameter values: $r_{max}=0.7$ and $N=100$.} 
\label{fig21}
\end{figure}

\begin{figure}[h!]
\begin{center}
\includegraphics[width=13cm]{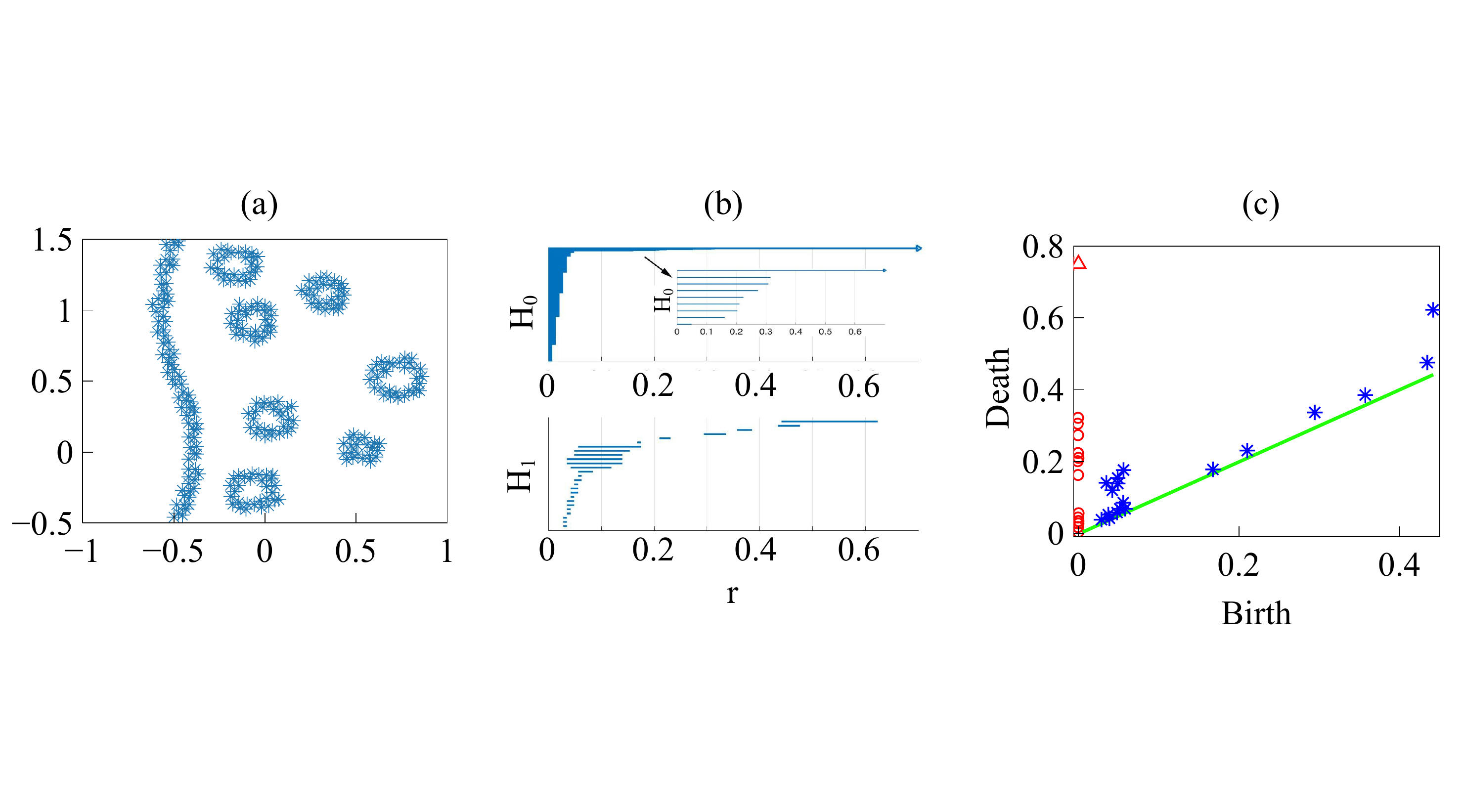} 
\end{center}
\vskip-1.5cm
\caption{Same as Fig.~\ref{fig20} except that there are one interface and seven islands.  
Parameter values: $r_{max}=0.7$ and $N=100$.}
\label{fig22}
\end{figure}

The {\em Vietoris-Rips filtration} (VRF) constructed in the main text provides an example. For each value of a scale proximity parameter $r>0$ and a given set of points $X$, we form a simplicial complex $VR(X,r)=S_r$ by finding all $k$-simplices such that all pairwise distances between their points are smaller than $r$. The simplicial complex $S_r$ comprises finitely many simplices such that (i) every nonempty subset of a simplex in $S_r$ is also in $S_r$, and (ii) two $k$-simplices in $S_r$ are either disjoint or intersect in a lower dimensional simplex. Clearly, if $r_1\leq r_2$, then $S_{r_1}\subset S_{r_2}$.  In $S_r$, 0-simplices are the data points, 1-simplices are edges, connections between two data points, 2-simplices are triangles formed by joining 3 data points through their edges, 3-simplices are tetrahedra, and we obtain more complicated structures for higher dimensional simplices. Fig.~\ref{fig19} shows a simple example of Vietoris-Rips complex, its barcodes for $\mathsf{H}_0$ and $\mathsf{H}_1$, and Betti numbers for different values of the proximity parameter $r$.  Panels (d)-(g) visualize the filtration process: For a grid of values of the filtration distance parameter $r$ we depict balls centered at points with radius $r$ and count the components formed. Topological features that persist on wide intervals of $r$ characterize the simplicial complexes of the dataset. To visualize persistent homology, we plot the {\em barcodes} and {\em persistence diagrams}. The barcode of a homology $\mathsf{H}_k$ depicts each class by its corresponding {\em Betti intervals} $(r_b,r_d)$. Initially we have one per point, represented as a bar in the top panel of Fig.~\ref{fig19}(b). As components merge, the number of bars diminishes. For panel (f) we have two components, represented by the two top  bars in panel $\mathsf{H}_0$ (b). For panel (g) we have one component represented by the top bar. The arrow means that this component persists for larger $r$ values. Similarly, the largest bar in panel $\mathsf{H}_1$ (b) represents the dominant hole, observed in panels  $(f)$-$(g)$. This bar, and hole, correspond to the circle in panel (c) placed furthest from the diagonal. The two small bars correspond to the circles in panel (c) which are closest to the  diagonal. As seen in panel (f), they form and disappear as components merge during the filtration process. In {\em persistence diagrams}, for the selected equally spaced grid of values of $r$, we represent each bar in the barcode by a point $(r_b,r_d)$ in the Cartesian plane. A point $(x,y)$ of the persistence diagram with multiplicity $m$ represents $m$ features that all appear for the first time at scale $x$ and disappear at scale $y$. The height of a point over the diagonal, $(y-x)$, gives the length of the corresponding bar in the barcode and is called the {\em persistence} of the feature. In addition to the off-diagonal points, the persistence diagram also contains each diagonal point, $(x, x)$, counted with infinite multiplicity. These additional points are needed for stability (discussed below) and make the cardinality of every persistence diagram infinite, even if the number of off-diagonal points is finite. Points near the diagonal are inferred to be noise while points further from the diagonal are considered topological signal \cite{top15}. Coloring differently different homologies ($\mathsf{H}_0$, $\mathsf{H}_1$, etc) we can accumulate plenty of topological information in one 2D persistence diagram. For example,
\begin{itemize}
\item $\mathsf{b}_0$ gives the number of components for a filtration value $r$, thereby providing the number of clusters in $\mathsf{H}_0$ for that $r$. We can use this information and the knowledge of the distance, to find out which points belong to which cluster. All the points of a  cluster are connected in a simplex. Thus, we have a clustering strategy.
\item Similarly, $\mathsf{b}_1$ gives the number of $1$-dimensional holes in $\mathsf{H}_1$ for a given value of the proximity parameter $r$. These holes may be inherent to the shape of a cluster or appear when basic clusters connect to more distant ones. See Figs.~\ref{fig20} (one island), \ref{fig21} (two islands) and \ref{fig22} (seven islands). Thus, $\mathsf{H}_1$ contains information on both the structure of basic clusters and their relative arrangements. The accompanying barcodes may characterize interfaces or data sets.
\end{itemize}

Next, we consider the homology of point clouds defining an interface between two populations, see Figs.~\ref{fig20}-\ref{fig22}.

Panels (d)-(f) in Figure \ref{fig20} visualize the filtration process: For a grid of values of the filtration distance parameter $r$, we depict balls centered at points with radius $r$ and count the components (clusters) formed. Initially we have as many clusters as points. As the filtration parameter $r$ increases, clusters merge and their number is reduced to two: the continuous front and the detached island in Fig.~\ref{fig20}(d), which are represented by the two top bars in $\mathsf{H}_0$ of Fig.~\ref{fig20}(b) and circles in Fig.~\ref{fig20}(c). Fig.~\ref{fig20}(e) exhibits only one cluster represented by the top bar in Fig.~\ref{fig20}(b), which persists for larger values of $r$ and is represented by an arrow. Similarly, the largest bar in $\mathsf{H}_1$ of Fig.~\ref{fig20}(b) represents the dominant hole defined by the island border in Fig.~\ref{fig20}(d). This bar, and island, corresponds to the asterisk in Fig.~\ref{fig20}(c) placed furthest from the diagonal. The small bar corresponds to the asterisk in Fig.~\ref{fig20}(c) which is closest to the diagonal. As seen in Fig.~\ref{fig20}(e), the small bar forms and disappears as holes form when  clusters merge during the filtration process.

Figs.~\ref{fig21} and \ref{fig22} can be similarly interpreted. Figs.~\ref{fig21}(d)-(f) visualize the filtration process. Initially we have one cluster per point, represented as a bar in the top panel of Fig.~\ref{fig21}(b). As the filtration parameter $r$ increases, clusters merge and the number of bars diminishes. Fig.~\ref{fig21}(d) exhibits three components, represented by the three top bars in $\mathsf{H}_0$ of Fig.~\ref{fig21}(b) and circles in Fig.~\ref{fig21}(c). There are two clusters in Fig.~\ref{fig21}(e) represented by the two top bars in Fig.~\ref{fig21}(b). Similarly, the two largest bars in $\mathsf{H}_1$ of Fig.~\ref{fig21}(b) represent the two dominant holes defined by the island border in Fig.~\ref{fig21}(d). These bars, and islands, correspond to the two asterisks in Fig.~\ref{fig21}(c) furthest from the diagonal. The small bar corresponds to the circle in Fig.~\ref{fig21}(c) which is closest to the diagonal. As seen in Fig.~\ref{fig21}(e), it forms and disappears as holes form when clusters merge during the filtration process.

\begin{figure}[!h]
\begin{center}
\includegraphics[width=14cm]{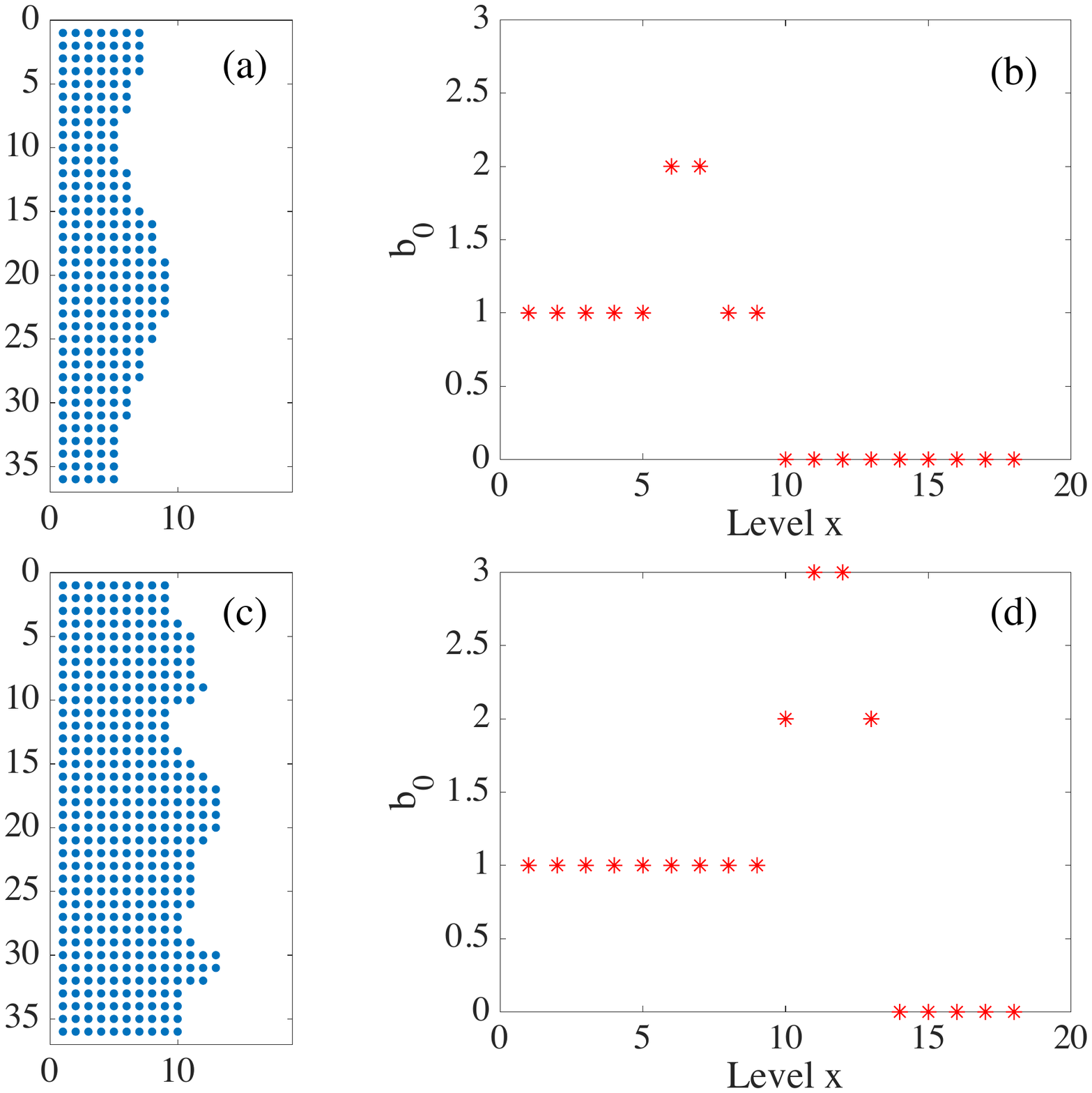} 
\end{center}
\caption{Interfaces of different roughness and Betti numbers $\mathsf{b}_0(r_{max})$ for the slices $x=c$ of the displayed  point clouds, as $c$ increases. We see how the variation in $\mathsf{b}_0$ gives an idea of the interface roughness, at the scale $r_{max}=1$ in this case. }
\label{fig23}
\end{figure}

\begin{figure}[!h]
\begin{center}
\includegraphics[width=14cm,angle=0]{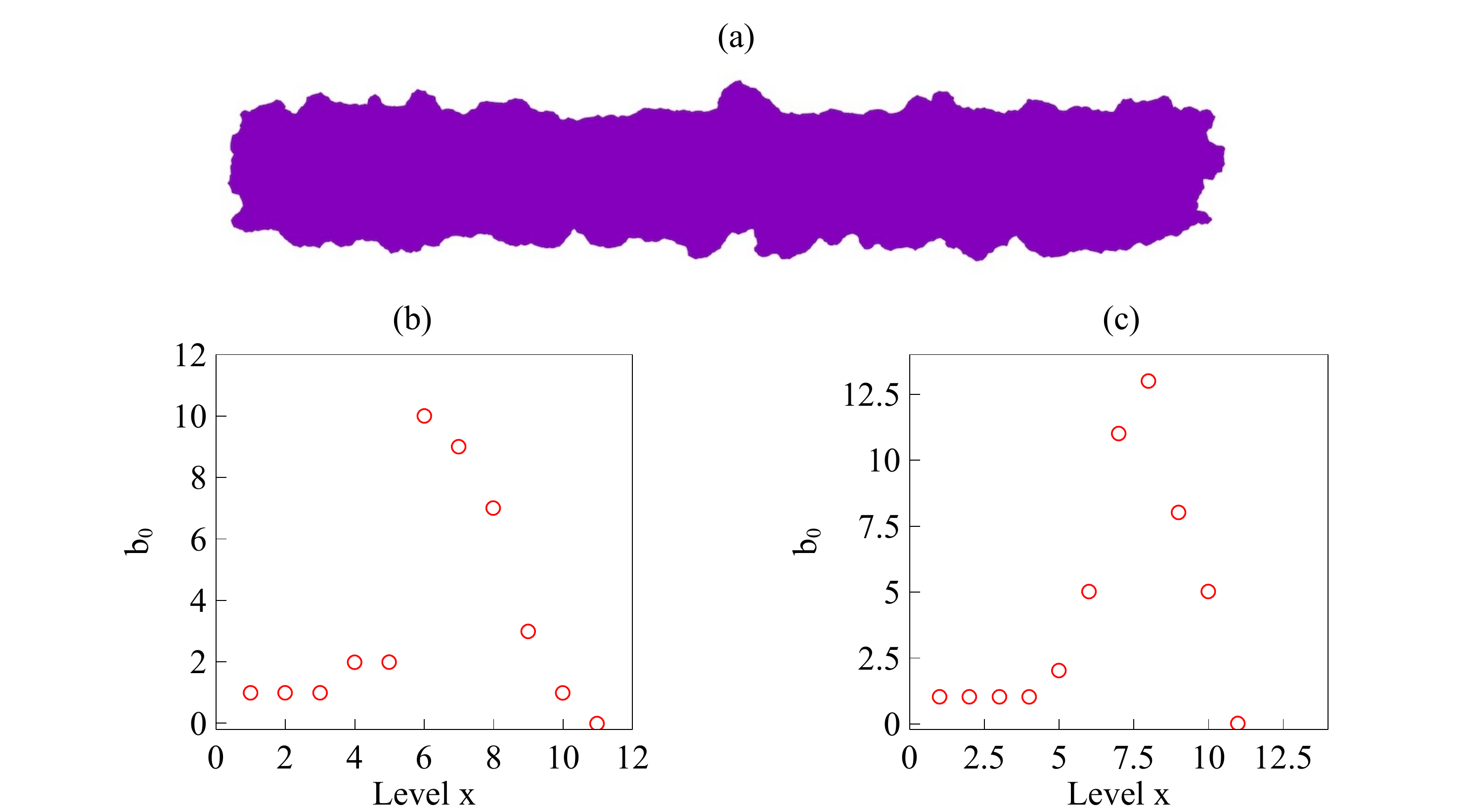} 
\end{center}
\caption{For the expanding strip in (a), we get the Betti numbers $\mathsf{b}_0$ for the upper front (b), and for the lower one (c). The lower front is rougher (larger $\mathsf{b}_0$) but its 
fingers are shorter ($\mathsf{b}_0$ decays faster to one and zero), whereas the upper front has a dominant persistent finger. We have set $r_{max}=1\!$ mm in the scale of the image Fig.~\ref{fig4}. }
\label{fig24}
\end{figure}

\begin{figure}[!h]
\begin{center}
\includegraphics[width=14cm,angle=0]{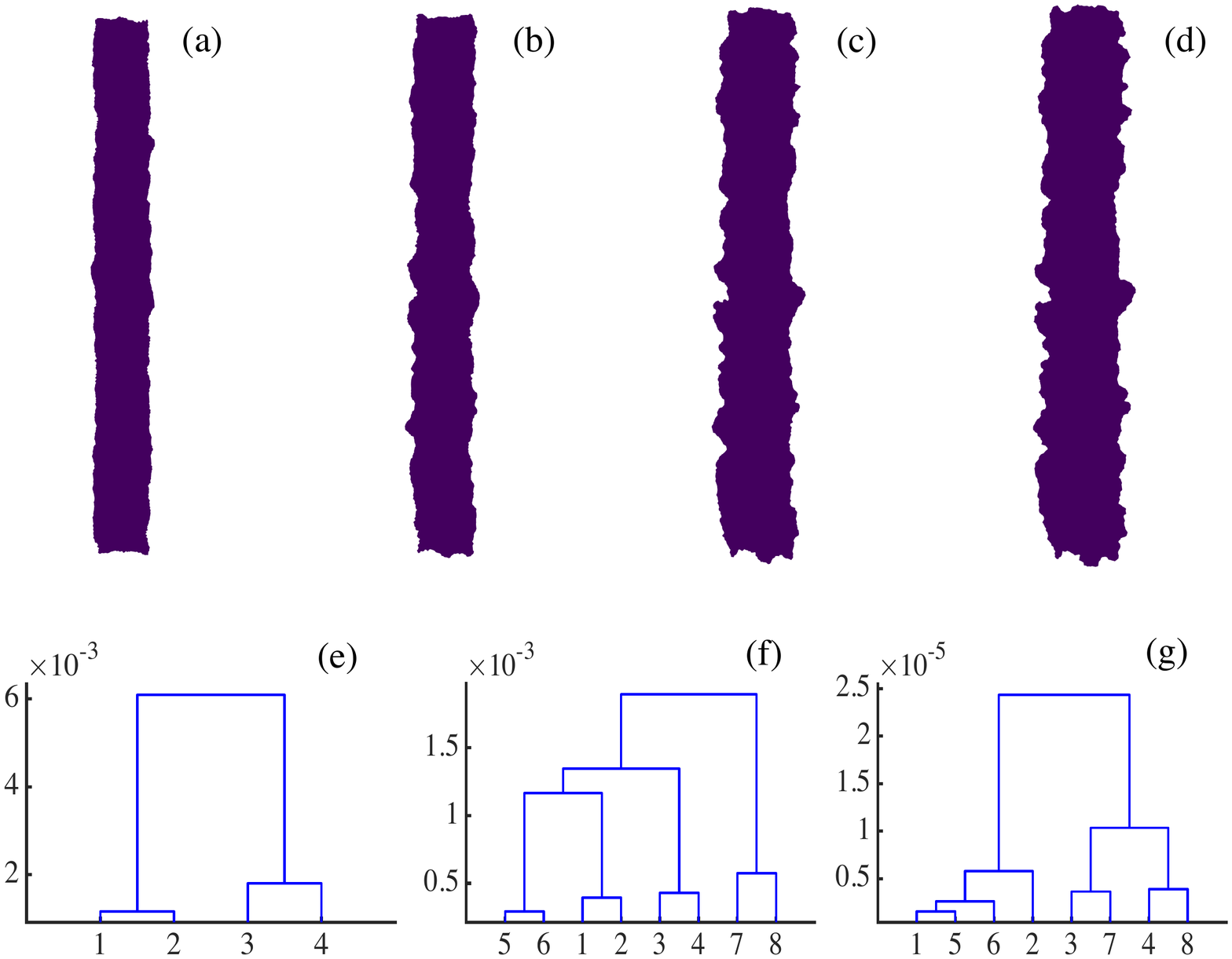}
\end{center}
\caption{(a)-(d) Consecutive snapshots of the evolution of the spreading 
configuration in Fig.~\ref{fig4}. (e)-(g) Dendrograms for hierarchical clustering constructed using Wasserstein distances $W_{1,\infty}$ between the four snapshots. We distinguish between overall snapshots (numbered 1 to 4)  in Panel (e), and half snapshots representing right (numbered 1 to 4) and left (numbered 5 to 8) moving interfaces in Panels (f) and (g). (e) The smoother overall snapshots 1 and 2 (corresponding to Panels (a) and (b)) are clustered together and, likewise, the rougher overall snapshots 3 and 4 of Panels (c) and (d). (f) The Wasserstein distance clusters together successive pairs of left and right fronts. (g) Dendrograms using the Betti number profiles $\mathsf{b}_0(r_{max})$ (analogous to those in Fig. \ref{fig24}) calculated for the left and right interfaces of each snapshot. Note that the interfaces of the first two snapshots are clustered together because they have similar roughness levels. }
\label{fig25}
\end{figure}

Fig.~\ref{fig22} exhibits an increased number of islands and it is described in the same manner. The main eight components correspond to one interface and seven islands. They are represented by the top bars in $\mathsf{H}_0$ of Fig.~\ref{fig22}(b) and circles in Fig.~\ref{fig22}(c). Seven bars represent the seven dominant islands, associated to the seven largest bars in $\mathsf{H}_1$ of Fig.~\ref{fig22}(b) and asterisks in Fig.~\ref{fig22}(c). The  remaining bars  represent gaps in the simplicial structure formed during the filtration process. The largest one represent a late hole appearing due to the fact that some islands are far from the main interface front, and corresponds to the final asterisk distant from the diagonal.

\paragraph{Tracking moving interfaces by tracking slices.} When the interface is simply connected, homology studies of the boundary points, or all the population points in the plane, hardly give information on its roughness. Instead we may study the evolution of the $\mathsf{H}_0$ homology of slices $x=c$ as $c$ varies, see Figs.~\ref{fig23} and \ref{fig24}. We choose as $r_{max}$ the degree of roughness we want to capture, the `scale' at which we wish to `resolve'. Consider the two dimensional region occupied by cells (magenta patch) in Figure \ref{fig24}(a). We build a square mesh of step smaller than $r_{max}$ and consider the points that are inside the occupied region. We define a matrix on the mesh $M(i,c)$, equal to one at points inside the patch, and zero outside. For each slice $x=c$, the points at which $M(i,c)=1$ define a point cloud, we evaluate the zero homology of that cloud using as maximum filtration value $r_{max}.$ The variation of the Betti number $\mathsf{b}_0(r_{max},c)$ with $c$ measures how irregular the front is: the larger $\mathsf{b}_0$ is, the rougher the interface. As shown in Fig.~\ref{fig24} corresponding to the spread assay of Fig.~\ref{fig4}(a), the rougher lower front has a larger maximal Betti number $\mathsf{b}_0$ than the upper front; cf Fig.~\ref{fig24}(c) versus Fig.~\ref{fig24}(b). However, the bigger and more persistent fingers of the upper front cause $\mathsf{b}_0$ to decay more slowly than the corresponding Betti number of the lower front. The fingers of the latter are smaller in size, cf Fig.~\ref{fig24}(a). Fig.~\ref{fig23} visualizes the idea on fragments of this configuration. This complementary slice by slice study gives qualitative information on the shape. This strategy would allow to study the time evolution of 2D interfaces comparing the information obtained for each time, and comparing the effect of different controlling parameters on each population, as done in the main text for mixing interfaces.  

Fig.~\ref{fig25} quantifies differences between configurations by means of distances between computational images \cite{liu}. To compare images and shapes, we first have to define measures $\rho^j(x)$, $x\in\Omega$, over grids of images $j$ ($j=0,1,2,\ldots$). Given two images, $j=0$ and 1, with measures $\rho^0(x)$ and $\rho^1(x)$, we define their {\em Wasserstein-1 distance} as a particular type of optimal transport distance~\cite{liu},
\begin{eqnarray*}  
&&W_{1,p}(\rho^0, \rho^1) =  \inf_{\pi: \Omega \times \Omega \to [0,\infty)} 
\int_{x, y \in \Omega} \| x - y \|_p\, \pi(x,y)\, dx dy, \quad 1 \leq p \leq \infty, \\
&&\mbox{\rm subject to}
\int_{y \in \Omega} \pi(x,y)\, dy = \rho^0(x) \,\, (x \in \Omega),\quad
\int_{x \in \Omega} \pi(x,y)\, dx = \rho^1(y) \,\,  (y \in \Omega),\quad
\pi(x,y) \geq 0 \,\, (x,y \in \Omega).  
\end{eqnarray*} 
Here $\pi(x,y)$ are probability measures over $\Omega\times\Omega$ whose marginals are $\rho^0(x)$ and $\rho^1(y)$. For images simply composed of points, which is the case of persistence diagrams, $\rho^0$ and $\rho^1$ are point measures and this definition recovers Eq.~\eqref{eq14}. Fig.~\ref{fig25}(e)-(g) uses $W_{1,\infty}$ distances between the four snapshots depicted in panels (a)-(d) to construct dendrograms for agglomerative hierarchical clustering following Ward's method \cite{agnes}. In Fig.~\ref{fig25}(e), we consider the four snapshots of Figs.~\ref{fig25}(a)-(d): the smoother first two snapshots  are clustered together and so are the last two rougher snapshots. Fig.~\ref{fig25}(f) does the same considering separately the left and right interfaces in each of the snapshots of Panels (a) to (d). In Fig.~\ref{fig25}(g), we cluster the $\mathsf{b}_0(r_{max})$ profiles obtained for the four right and left fronts. Dendrograms enable us to do this. Enforcing cutoffs on the inconsistency coefficients we obtain a natural division in three clusters $\{1,2,5,6\}$, $\{3,7\}$ and $\{4,8\}$. Fronts of similar roughness are clustered together. The same clusters are obtained enforcing
cutoffs on distances, that is, cutting the dendrogram at a height that defines
three clusters or by K-means with three clusters. Thus, the results are robust.
Cluster analysis for Fig.~\ref{fig25}(f) shows a higher variability depending on the method 
employed. Enforcing cutoffs on the inconsistency coefficients we obtain a 
natural division in four clusters $\{1,2\}$, $\{3,4\}$, $\{5,6\}$ and $\{7,8\}$. If we
seek for a smaller number of clusters through distance cutoffs or K-means,
the results vary. Other clustering methods, such as single linkage, yield lower cophenetic correlation coeficients, which means that Ward's clustering represents these data slightly better.  

\section*{Supporting Information} 
 {\textbf S1 Supporting movies and data files for TDA with a README text file. }

\end{document}